\documentclass[sort&compress,a4paper,oneside,3p,12pt]{melsarticle}
\usepackage{verbatim}
\usepackage{amssymb}
\usepackage{amsmath,amssymb,amsfonts,graphicx,float,multicol,fleqn}
\usepackage{multirow}
\usepackage{amssymb}
\usepackage{amsmath,amsthm}
\usepackage{amsfonts}
\usepackage{graphicx}
\usepackage{graphics}
\input{psfig.sty}
\numberwithin{equation}{section}
\newtheorem{theorem}{Theorem}[section]

\begin{document}

\begin{frontmatter}

%% Title, authors and addresses

%% use the tnoteref command within \title for footnotes;
%% use the tnotetext command for the associated footnote;
%% use the fnref command within \author or \address for footnotes;
%% use the fntext command for the associated footnote;
%% use the corref command within \author for corresponding author footnotes;
%% use the cortext command for the associated footnote;
%% use the ead command for the email address,
%% and the form \ead[url] for the home page:
%%
%% \title{Title\tnoteref{label1}}
%% \tnotetext[label1]{}
%% \author{Name\corref{cor1}\fnref{label2}}
%% \ead{email address}
%% \ead[url]{home page}
%% \fntext[label2]{}
%% \cortext[cor1]{}
%% \address{Address\fnref{label3}}
%% \fntext[label3]{}

\title{An approximation algorithm for the solution of the nonlinear Lane-Emden type equations arising in astrophysics using Hermite functions collocation method}

%% use optional labels to link authors explicitly to addresses:
%% \author[label1,label2]{<author name>}
%% \address[label1]{<address>}
%% \address[label2]{<address>}

\author[a,b]{K. Parand}
\ead{k\_parand@sbu.ac.ir}
\fntext[a]{Member of research group of Scientific Computing.}
\author[c]{Mehdi Dehghan\corref{cor}}
\cortext[cor]{Corresponding author. Fax: (9821) 66497030.}
\ead{mdehghan.aut@gmail.com}
\author[b]{A.R. Rezaei}
\ead{alireza.rz@gmail.com}
\author[b]{S.M. Ghaderi}
\ead{ghaderi@khayam.ut.ac.ir}
\address[b]{Department of Computer Sciences, Shahid Beheshti University, G.C., Tehran, Iran}
\address[c]{Department of Applied Mathematics, Faculty of Mathematics and Computer Science, Amirkabir University of Technology, No.424, Hafez Avenue, Tehran 15914, Iran}
%%%%%%%%%%%%%%%%%%%%%%%%%%%%%%%%%%%%%%%%%%%%%%%%%%%%%%%%

\begin{abstract}
In this paper we propose a collocation method for solving some well-known classes of Lane-Emden type equations which are nonlinear ordinary differential equations on the semi-infinite domain. They are categorized as singular initial value problems. The proposed approach is based on a Hermite function collocation (HFC) method. To illustrate the reliability of the method, some special cases of the equations are solved as test examples. The new method reduces the solution of a problem to the solution of a system of algebraic equations. Hermite functions have prefect properties that make them useful to achieve this goal. We compare the present work with some well-known results and show that the new method is efficient and applicable.
\end{abstract}


\end{frontmatter}
%*********************************Introduction***********************************************

\section{Introduction}
Many problems in science and engineering arise in unbounded domains.
Different spectral methods have been proposed for solving problems on unbounded domains.
The most common method is through the use of polynomials that are orthogonal over unbounded domains, such as the Hermite spectral and the Laguerre spectral methods \cite{Coulaud,Funaro.Kavian,Funaro.Appl. Numer. Math.1990,Guo.Math. Comp.1999,Guo.num2000,Maday,Shen,Siyyam}.

Guo \cite{Guo.J. Math. Anal. Appl.1998,Guo.com2000,Guo.J. Math. Anal. Appl.2000} proposed a method that proceeds by mapping the original problem in an unbounded domain to a problem in a bounded domain, and then using suitable Jacobi polynomials to approximate the resulting problems.

Another approach is replacing the infinite domain with $[-L, L]$ and the semi-infinite interval with $[0, L]$ by choosing $L$, sufficiently large. This method is named as the domain truncation \cite{BoydBook}.

Another effective direct approach for solving such problems is based on rational approximations.
Christov \cite{Christov.SIAM J. Appl. Math.1982} and Boyd \cite{Boyd.J. Comput. Phys.1987(69),Boyd1987} developed some spectral methods on unbounded intervals by using mutually orthogonal systems of rational functions.
Boyd \cite{Boyd1987} defined a new spectral basis, named rational Chebyshev functions on the semi-infinite interval, by mapping it to the Chebyshev polynomials.
Guo et al. \cite{Guo.sci2000} introduced a new set of rational Legendre functions which are mutually orthogonal in $L^2(0,+\infty)$.
They applied a spectral scheme using the rational Legendre functions for solving the Korteweg-de Vries equation
on the half line.
Boyd et al. \cite{Boyd2003} applied pseudospectral methods on a semi-infinite interval and compared the rational Chebyshev, Laguerre and the mapped Fourier sine methods.

Authors of~\cite{Parand.Appl. Math. Comput.2004,Parand.Int. J. Comput. Math.2004,Parand.Phys. Scripta2004,Parand.Phys.let.A,Parand.CAM,Parand.JCP} applied the spectral method to solve the nonlinear ordinary differential equations on semi-infinite intervals. Their approach is based on the rational Tau and collocation methods.

Lane-Emden type equations are nonlinear ordinary differential equations on semi-infinite domain. They are categorized as singular initial value problems.
These equations describe the temperature variation of a spherical gas cloud under the mutual attraction of its molecules and subject to the laws of classical thermodynamics.
The polytropic theory of stars essentially follows out of thermodynamic considerations, that deals with the issue of energy transport, through the transfer of material between different levels of the star. These equations are one of the basic equations in the theory of stellar structure and has been the focus of many studies \cite{Bender.J. Math. Phys.1989,Mandelzweig,Shawagfeh,Wazwaz.Appl,Wazwaz.Appl. Math. Comput.2006,Liao.Appl. Math. Comput.2003,He.Appl. Math. Comput.2003,Ramos.Comp. Phys. Commmun.2003,Ramos.Appl. Math. Comput.2005,Ramos.Chaos Soliton. Frac.2008,Ramos.Chaos Soliton. Frac.2009,Yousefi,Bataineh.Commun. Nonlinear Sci. Numer. Simul.2008,ChowdhuryHashim,Aslanov.Phys. Lett. A2008,Dehghan.New Astron.2008,Agarwal.ORegan2007}.

We simply begin with the Poisson equation and the condition for hydrostatic equilibrium \cite{Aslanov.Phys. Lett. A2008,Dehghan.New Astron.2008,Agarwal.ORegan2007,Chandrasekhar}:
\begin{eqnarray}\nonumber
\frac{\mathrm{d}P}{\mathrm{d}r}=-\rho\frac{GM(r)}{r^2},
\end{eqnarray}
\begin{eqnarray}\nonumber
\frac{\mathrm{d}M(r)}{\mathrm{d}r}=4\pi \rho{r^2},
\end{eqnarray}
where $G$ is the gravitational constant, $P$ is the pressure, $M(r)$ is the mass of a star at a certain radius $r$, and $\rho$ is the density, at a distance $r$ from the center of a spherical star \cite{Chandrasekhar}.
The combination of these equations yields the following equation, which as should be noted, is an equivalent form of the Poisson equation \cite{Aslanov.Phys. Lett. A2008,Dehghan.New Astron.2008,Chandrasekhar,Agarwal.ORegan2007}:
\begin{eqnarray}\nonumber
\frac{1}{r^2}\frac{\mathrm{d}}{\mathrm{d}r}\left(\frac{r^2}{\rho}\frac{\mathrm{d}P}{\mathrm{d}r}\right)=-4\pi G\rho.
\end{eqnarray}
From these equations one can obtain the Lane-Emden equation through the simple assumption that the pressure is
simply related to the density, while remaining independent of the temperature.
We already know that in the case of a degenerate electron gas, the pressure and density are $\rho\thicksim P^\frac{3}{5}$, assuming that such a relation exists for other states of the star, we are led to consider a relation of the following form \cite{Chandrasekhar}:
\begin{eqnarray}\nonumber
P=K\rho^{1+\frac{1}{m}},
\end{eqnarray}
where $K$ and $m$ are constants, at this point it is important to note that $m$ is the polytropic index which is related to the ratio of specific heats of the gas comprising the star.
Based upon these assumptions we can insert this relation into our first equation for the hydrostatic equilibrium condition and from this equation \cite{Aslanov.Phys. Lett. A2008,Dehghan.New Astron.2008,Chandrasekhar,Agarwal.ORegan2007} we have
\begin{eqnarray}\nonumber
\left[\frac{K(m+1)}{4\pi G}\lambda^{\frac{1}{m}-1}\right] \frac{1}{r^2}\frac{\mathrm{d}}{\mathrm{d}r}\left(r^2\frac{\mathrm{d}y}{\mathrm{d}r}\right)=-y^m,
\end{eqnarray}
where the additional alteration to the expression for density has been inserted with $\lambda$ representing
the central density of the star and $y$ that of a related dimensionless quantity that are both related to $\rho$ through the following relation \cite{Aslanov.Phys. Lett. A2008,Dehghan.New Astron.2008,Chandrasekhar,Agarwal.ORegan2007}
\begin{eqnarray}\nonumber
\rho=\lambda y^m.
\end{eqnarray}
Additionally, if place this result into the Poisson equation, we obtain a differential equation for the mass, with a dependance upon the polytropic index $m$.
Though the differential equation is seemingly difficult to solve, this problem can be partially alleviated by the introduction of an additional dimensionless variable $x$, given by the following:
\begin{eqnarray}\nonumber
r=ax,
\end{eqnarray}
\begin{eqnarray}\nonumber
a=\left[\frac{K(m+1)}{4\pi G}\lambda^{\frac{1}{m}-1}\right]^\frac{1}{2}.
\end{eqnarray}
Inserting these relations into our previous equations we obtain the famous form of the Lane-Emden equations, given
in the following:
\begin{eqnarray}\nonumber
\frac{1}{x^2}\frac{\mathrm{d}}{\mathrm{d}x}\left(x^2\frac{\mathrm{d}y}{\mathrm{d}x}\right)=-y^m.
\end{eqnarray}
Taking these simple relations we will have the standard Lane-Emden equation with $g(y)=y^m$ \cite{Aslanov.Phys. Lett. A2008,Dehghan.New Astron.2008,Chandrasekhar,Agarwal.ORegan2007},
\begin{equation}\label{Eq.Lane-Emden.asli equation}
y''+\frac{2}{x}y'+y^m=0,\qquad x>0.
\end{equation}
At this point it is also important to introduce the boundary conditions which are based upon the following boundary conditions for hydrostatic equilibrium and normalization consideration of the newly introduced quantities $x$ and $y$. What follows for $r=0$ is
\begin{equation}\label{Eq.Lane-Emdenboun equation}
r=0 \rightarrow x=0,\quad       \rho=\lambda\rightarrow y(0)=1.
\end{equation}
As a result an additional condition must be introduced in order to maintain the condition of
Eq. (\ref{Eq.Lane-Emdenboun equation}) simultaneously:
\begin{eqnarray}\nonumber
y'(0)=0.
\end{eqnarray}
In other words, the boundary conditions are as follows
\begin{eqnarray}\nonumber
y(0)=1,\quad y'(0)=0.
\end{eqnarray}
The values of $m$ which are physically interesting,
lie in the interval [0,5]. The main difficulty in the analysis of this type of equation is the singularity behavior occurring at $x=0$.\\
Exact soloutions for Eq. (\ref{Eq.Lane-Emden.asli equation}) are known only for $m=0,1$ and $5$. For other values of $m$ the standard Lane-Emden equation is to be integrated numerically.
Thus we decided to present a new and efficient technique to solve it numerically.

This paper is arranged as follows:\\
In Section \ref{MethodsusedtoLEE} we survey several methods that have been used to solve Lane-Emden type equations.
In Section \ref{HFCmethod}, the properties of Hermite functions and the way to construct the collocation technique for this type of equation are described. In Section \ref{Applications} the proposed method is applied to some types of Lane-Emden equations, and a comparison is made with
the existing analytic or exact solutions that were reported in other published works in the literature. Finally we give a brief conclusion in the last section.

%--------------------------------------------------------------------------------
\section{Methods have been proposed to solve Lane-Emden equations}\label{MethodsusedtoLEE}
%*********************************Solved Lane-Emden******************************************
Recently, many analytical methods have been used to solve Lane-Emden equations, the main difficulty arises in the
singularity of the equations at $x = 0$. Currently, most techniques which were used in handling the Lane-Emden-type problems are based on either series solutions or perturbation techniques.\\
Bender et al. \cite{Bender.J. Math. Phys.1989} proposed a new perturbation technique based on an
artificial parameter $\delta$, the method is often called $\delta$-method.

Mandelzweig et al. \cite{Mandelzweig} used the quasilinearization approach to solve the standard Lane-Emden equation.
This method approximates the solution of a nonlinear differential equation by treating the nonlinear terms as a perturbation about the linear ones, and unlike perturbation theories is not based on the existence of some small parameters.

Shawagfeh \cite{Shawagfeh} applied a nonperturbative approximate analytical solution for the Lane-Emden
equation using the Adomian decomposition method.
His solution was in the form of a power series.
He used Pad\'{e} approximants method~\cite{ShakourifarCsf09,DehghanAmc07} to accelerate the convergence of the power series.

In \cite{Wazwaz.Appl}, Wazwaz employed the Adomian decomposition method~\cite{DehghanScripta08,DehghanMpe06} with an alternate framework designed to overcome the difficulty of the singular point.
It was applied to the differential equations of Lane-Emden type.
Further author of~\cite{Wazwaz.Appl. Math. Comput.2006} used the modified decomposition method for solving the analytical treatment of nonlinear differential equations such as the Lane-Emden equation.

Liao \cite{Liao.Appl. Math. Comput.2003} provided an analytical algorithm for Lane-Emden type equations.
%%%%
This algorithm logically contains the well-known Adomian decomposition method.
%%%%
Different from all other analytical techniques, this algorithm itself provides us with a convenient way to adjust convergence regions even without Pad\'{e} technique.

J.-H He~\cite{He.Appl. Math. Comput.2003} employed Ritz's method to obtain an analytical solution
of the problem. By the semi-inverse method, a variational principle is obtained for the Lane-Emden
equation.

Parand et al. \cite{Parand.Int. J. Comput. Math.2004,Parand.Phys. Scripta2004,Parand.JCP} presented two numerical techniques to solve higher ordinary differential equations such as Lane-Emden.
Their approach was based on the rational Chebyshev  and rational Legendre Tau methods.

Ramos \cite{Ramos.Comp. Phys. Commmun.2003,Ramos.Appl. Math. Comput.2005,Ramos.Chaos Soliton. Frac.2008,Ramos.Chaos Soliton. Frac.2009} solved Lane-Emden equations through different methods.
Author of~\cite{Ramos.Appl. Math. Comput.2005} presented the linearization method for singular initial-value problems in second-order ordinary differential equations such as Lane-Emden.
These methods result in linear constant-coefficients ordinary differential equations which can be integrated analytically, thus yielding piecewise analytical solutions and globally smooth solutions.
Later this author~\cite{Ramos.Chaos Soliton. Frac.2009} developed piecewise-adaptive decomposition methods for the solution of nonlinear ordinary differential equations. In \cite{Ramos.Chaos Soliton. Frac.2008}, series solutions of the Lane-Emden type equation have been obtained by writing this equation as a Volterra integral
equation and assuming that the nonlinearities are sufficiently differentiable. These series solutions have been obtained by either working with the original differential equation or transforming it into an ordinary differential equation that does not contain the first-order derivatives. Series solutions to the Lane-Emden type equation have also been obtained by working directly on the original differential equation or transforming it into a simpler one.

Yousefi \cite{Yousefi} presented a numerical method for solving the Lane-Emden equations. He converted Lane-Emden equations to integral equations, using integral operator, and then he applied Legendre wavelet approximations.

Bataineh et al. \cite{Bataineh.Commun. Nonlinear Sci. Numer. Simul.2008} presented an algorithm based on
homotopy analysis method (HAM)~\cite{DehghanNmpde10} to obtain the approximate analytical
solutions of the singular IVPs of the Emden-Fowler type equation.

In \cite{ChowdhuryHashim}, Chowdhury et al. presented an algorithm based on the homotopy-perturbation method (HPM)
~\cite{DehghanMcm08,DehghanJpm08,DehghanNonrwa09} to solve singular IVPs of time-independent equations.

Aslanov \cite{Aslanov.Phys. Lett. A2008} introduced a further development in the Adomian decomposition method to overcome the difficulty at the singular point of non-homogeneous, linear and non-linear Lane-Emden-like equations.

Dehghan and Shakeri \cite{Dehghan.New Astron.2008} applied an exponential transformation to the Lane-Emden type equations to overcome the difficulty of a singular point at $x = 0$ and solved the resulting nonsingular problem by the variational iteration method~\cite{DehghanCam07,SaadatmandiCsf09}.

%%%%%%%%%%%%%%%%%
Yildirim et al. \cite{Yildirim} presented approximate-exact solutions of a class of Lane-Emden type singular IVPs
problems, by the variational iteration method.
%%%%%%%%%%%%%%%%

Marzban et al. \cite{Marzban} used a method based upon hybrid function approximations. They used the properties of hybrid of block-pulse functions and Lagrange interpolating polynomials together for solving the nonlinear second-order initial value problems and the Lane-Emden equation.

Recently, Singh et al. \cite{Singh.Pandey2009} provided an efficient analytic algorithm for Lane-Emden type equations using modified homotopy analysis method, also they used some well-known Lane-Emden type equations as test examples.

We refer the interested reader to~\cite{Kara1,Kara2} for analysis of the Lane-Emden equation based on
the Lie symmetry approach.

%---------------------------------------------------------------------------------------------
\section{Hermite functions collocation method}\label{HFCmethod}
% ********************************************************************
%%\section{Solving differential boundary value problems defined in unbounded domains with Spectral methods}
Spectral methods have been successfully applied in the approximation of
boundary value problems defined in unbounded domains. For problems whose solutions are
sufficiently smooth, they exhibit exponential rates of convergence/spectral accuracy.
We can apply different spectral methods that are used to solve problems in the
semi-infinite domains. One of these approaches is using Laguerre and Hermite polynomials/functions \cite{Funaro.Kavian,Guo.Math. Comp.1999,Guo.num2000,Maday,Shen,Siyyam,GuoShenXu2003,Bao.Shen,Guo.Xu}. Guo \cite{Guo.num2000} suggested the Laguerre-Galerkin method for the Burgers' equation and Benjamin-Bona-Mahony (BBM) equation on a semi-infinite interval.
In \cite{Shen} Shen proposed spectral methods using Laguerre functions and analyzed the elliptic equations on regular unbounded domains.
Siyyam \cite{Siyyam} applied two numerical methods for solving differential equations using the Laguerre Tau method.
Maday et al. \cite{Maday} proposed a Laguerre type spectral method for solving partial
differential equations.
Funaro and Kavian \cite{Funaro.Kavian} considered some algorithms by using the Hermite functions. Recently Guo \cite{Guo.Math. Comp.1999} developed the spectral method by using Hermite polynomials. However it is not easy to perform the quadratures in unbounded domains, which are used in the Hermite spectral approximations. So the Hermite pseudospectral method is more preferable in actual calculations. Guo \cite{Guo.Xu} developed the Hermite pseudospectral method for the Burgers' equation on the whole line. Guo et al.~\cite{GuoShenXu2003} considered spectral and pseudospectral approximations using Hermite functions for partial differential equations (PDEs) on the whole line to approximate the Dirac equation. Bao and Shen \cite{Bao.Shen} proposed a generalized-Laguerre-Hermite pseudospectral method for computing symmetric and central vortex states in Bose-Einstein condensates (BECs) in three dimensions with cylindrical symmetry.

Collocation method~\cite{DehghanIjnsns06} has become increasingly popular for solving differential equations. Also they are very useful in providing highly accurate solutions to differential equations.
In this paper, we employ the Hermite functions collocation (denoted by HFC) method to solve some well-known singular forms of the Lane-Emden type initial value problems directly.
\subsection{Properties of Hermite functions}
In this section, we detail the properties of the Hermite functions that will be used to construct the HFC method.
First we note that the Hermite polynomials are generally not suitable in practice due to their wild asymptotic behavior at infinities \cite{ShenWang2008}.\\
Hermite polynomials can be written in direct formula as follows:
%--------------------------------------------------------------------------------

\begin{align}\nonumber
H_{n}(x)&\sim\frac{\Gamma(n+1)}{\Gamma(n/2+1)}e^{x^2/2}\cos{(\sqrt{2n+1}x-\frac{n\pi}{2})}\\\nonumber
&\sim n^{n/2}e^{x^2/2}\cos(\sqrt{2n+1}x-\frac{n\pi}{2}).
\end{align}
Hence, we shall consider the so called Hermite functions.
The normalized Hermite functions of degree $n$ is defined by
\begin{eqnarray}\nonumber
\widetilde{H}_{n}(x)=\frac{1}{\sqrt{2^n n!}}e^{-x^2/2}H_{n}(x), \quad n\geq 0, x\in \mathbb{R}.
\end{eqnarray}
Clearly, \{$\widetilde{H}_{n}$\} is an orthogonal system in $L^2(\mathbb{R})$,i.e.,
\begin{eqnarray}\nonumber
\int^{+\infty}_{-\infty}\widetilde{H}_{n}(x)\widetilde{H}_{m}(x)dx=\sqrt{\pi}\delta_{mn},
\end{eqnarray}
where $\delta_{nm}$ is the Kronecker delta function.
In contrast to the Hermite polynomials, the Hermite functions are well behaved with the decay property:
\begin{eqnarray}\nonumber
|\widetilde{H}_{n}(x)|\longrightarrow 0, \quad \text{as}\quad |x|\longrightarrow \infty,
\end{eqnarray}
and the asymptotic formula with large $n$ is
\begin{eqnarray}\nonumber
\widetilde{H}_{n}(x)\sim n^{-\frac{1}{4}}\cos(\sqrt{2n+1}x-\frac{n\pi}{2}).
\end{eqnarray}
The three-term recurrence relation of Hermite polynomials implies
\begin{eqnarray}\nonumber
&\widetilde{H}_{n+1}(x)=x\sqrt{\frac{2}{n+1}}\widetilde{H}_{n}(x)-\sqrt{\frac{n}{n+1}}\widetilde{H}_{n-1}(x),\quad n\geq{1},\\\nonumber
&\noindent\widetilde{H}_{0}(x)=e^{-x^2/2}, \quad \widetilde{H}_{1}(x)=\sqrt{2}xe^{-x^2/2}.
\end{eqnarray}
Using the recurrence relation of Hermite polynomials and the above formula lead to
\begin{align}\nonumber
\widetilde{H}'_{n}(x)&=\sqrt{2n}\widetilde{H}_{n-1}(x)-x\widetilde{H}_{n}(x)\\\nonumber
&=\sqrt{\frac{n}{2}}\widetilde{H}_{n-1}(x)-\sqrt{\frac{n+1}{2}}\widetilde{H}_{n+1}(x).
\end{align}
and this implies
\begin{eqnarray}\nonumber
\int_{\mathbb{R}}{\widetilde{H}'_n(x)\widetilde{H}'_m(x)dx}=
\begin{cases}
-\frac{\sqrt{n(n-1)\pi}}{2}, & m=n-2, \\
\sqrt{\pi}(n+\frac{1}{2}), & m=n,\\
-\frac{\sqrt{(n+1)(n+2)\pi}}{2}, & m=n+2, \\
0, & \text{otherwise}.
\end{cases}
\end{eqnarray}
Let us define
\begin{eqnarray}\nonumber
\widetilde{P}_N:=\{u:u=e^{-x^2/2}v,\forall v\in{P_N}\},
\end{eqnarray}
where $P_N$ is the set of all Hermite polynomials of degree at most $N$.\\
We now introduce the Gauss quadrature associated with the Hermite functions approach.\\
Let $\{x_j\}_{j=0}^{N}$ be the Hermite-Gauss nodes and define the weights
\begin{eqnarray}\nonumber
\widetilde{w}_j=\frac{\sqrt{\pi}}{(N+1)\widetilde{H}_{N}^{2}(x_j)}, \quad 0\leq j\leq N.
\end{eqnarray}
Then we have
\begin{eqnarray}\nonumber
\int_{\mathbb{R}}p(x)dx=\sum_{j=0}^{N}p(x_j)\widetilde{w}_j, \quad \forall p\in \widetilde{P}_{2N+1}.
\end{eqnarray}
For a more detailed discussion of these early developments see~\cite{ShenTangHighOrder,ShenTangWang}.
\subsection{Approximations by Hermite functions}
Let us define $\Lambda:=\{x|-\infty<x<\infty\}$
and
\begin{eqnarray}\nonumber
\mathcal{H}_N=span\{\widetilde{H}_0(x),\widetilde{H}_1(x),...,\widetilde{H}_ N(x)\}.
\end{eqnarray}
The $L^{2}{(\Lambda)}$-orthogonal projection $\tilde{\xi}_N:L^{2}{(\Lambda)}\longrightarrow \mathcal{H}_N$
is a mapping in a way that for any $v\in L^{2}{(\Lambda)}$,
\begin{eqnarray}\nonumber
<\tilde{\xi}_N{v}-v,\phi>=0, \quad \forall \phi\in \mathcal{H}_N,
\end{eqnarray}
or equivalently,
\begin{eqnarray}\nonumber
\tilde{\xi}_N{v(x)}=\sum_{l=0}^{N}\tilde{v}_{l}\widetilde{H}_l(x).
\end{eqnarray}
To obtain the convergence rate of Hermite functions we define the space $H_{A}^{r}(\Lambda)$ defined by
\begin{eqnarray}\nonumber
H^{r}_{A}{(\Lambda)}=\{v|v \text{ is measurable on } \Lambda \text{ and }{\|v\|}_{r,A}<\infty\},\nonumber
\end{eqnarray}
and equipped it with the norm $\|v\|_{r,A}=\|A^{r}v\|$. For any $r > 0$, the space $H^{r}_{A}{(\Lambda)}$
and its norm are defined by space interpolation. By induction, for any non-negative integer $r$ we can write
\begin{eqnarray}\nonumber
A^{r}v(x)=\sum_{k=0}^{r}(x^2+1)^{(r-k)/2}p_k(x){\partial}_{x}^{k}v(x),
\end{eqnarray}
where $p_k(x)$ are certain rational functions which are bounded uniformly on $\Lambda$. Thus we have
\begin{eqnarray}\nonumber
\|v\|_{r,A}\leq c \left(\sum_{k=0}^{r}\parallel(x^2+1)^{(r-k)/2}p_k(x)\partial_{x}^{k}v\parallel \right)^{1/2}.
\end{eqnarray}
\begin{theorem}\nonumber
For any $v \in H^{r}_{A}(\Lambda)$, $r\geq1$ and $0\leq \mu \leq r$, the following can be obtained:
\begin{equation}
\parallel{\tilde{\xi}}_{N}v-v{\parallel_{\mu}\leq cN^{1/3+(\mu-1)/2}\parallel v\parallel}_{r,A}.
\end{equation}
\end{theorem}
{\it Proof}. A complete proof is given by Guo et al. \cite{GuoShenXu2003}.
Also same theorems have been proved by Shen et al. \cite{ShenWang2008}.
%//////////////////////////////////////////////////////////////////////
\subsection{Hermite functions transform}
As mentioned before, Lane-Emden type equations are defined on the interval $(0,+\infty)$; but we know properties of Hermite functions are derived in the infinite domain $(-\infty,+\infty)$.
%%%%
Also we know approximations can be constructed for infinite, semi-infinite and finite intervals.
One of the approaches to construct approximations on the interval $(0,+\infty)$ which is used in
the current paper, is to use a mapping, that is a change of variable of the form
\begin{equation}\nonumber
\omega=\phi(z)=\ln(\sinh(kz)),
\end{equation}
where $k$ is a constant.\\
The basis functions on $(0,+\infty)$ are taken to be the transformed Hermite functions,
\begin{eqnarray}\nonumber
\widehat{H}_n(x)\equiv \widetilde{H}_n(x)\circ \phi(x)= \widetilde{H}_n(\phi(x)),
\end{eqnarray}
where $\widetilde{H}_n(x)\circ \phi(x)$ is defined by $\widetilde{H}_n(x)(\phi(x))$. The inverse map of $\omega=\phi(z)$ is
\begin{eqnarray}\label{inverseTransform}
z=\phi^{-1}(\omega)=\frac{1}{k}\ln (e^\omega +\sqrt{e^{2\omega}+1}).
\end{eqnarray}
Thus we may define the inverse images of the spaced nodes ${ {\{{x_j}}\}_{x_j=-\infty}^{x_j=+\infty} }$ as
\begin{eqnarray}\nonumber
\Gamma=\{\phi^{-1}(t): -\infty< t <+\infty\}=(0,+\infty),
\end{eqnarray}
and
\begin{eqnarray}\nonumber
\tilde{x}_j=\phi^{-1}(x_j)=\frac{1}{k}\ln (e^{x_j} +\sqrt{e^{2x_j}+1}),\quad j=0,1,2,...
\end{eqnarray}
Let $w(x)$ denotes a non-negative, integrable, real-valued function over the interval $\Gamma$.
We define
\begin{eqnarray}\nonumber
L^2_w(\Gamma)=\{v:\Gamma\rightarrow \mathbb{R}\mid v \textrm{is measurable and}\parallel v{\parallel}_w<\infty \},
\end{eqnarray}
where
\begin{eqnarray}\nonumber
\parallel v{\parallel}_w=\left(\int_{0}^\infty\mid v(x)\mid ^2w(x)\mathrm{d}x\right)^{\frac{1}{2}},
\end{eqnarray}
is the norm induced by the inner product of the space $L^2_w(\Gamma)$,
\begin{equation}\label{Eq.inner product definition}
<u,v>_w=\int_{0}^{\infty}u(x)v(x)w(x)\mathrm{d}x.
\end{equation}
Thus $\{\widehat{H}_n(x)\}_{n\in \mathbb{N}}$ denotes a system which is mutually orthogonal under (\ref{Eq.inner product definition}), i.e.,
\begin{eqnarray}\nonumber
< \widehat{H}_n(x),\widehat{H}_m(x)>_{w(x)}=\sqrt{\pi}\delta_{nm},
\end{eqnarray}
where $w(x)=\coth(x)$ and $\delta_{nm}$ is the Kronecker delta function. This system is complete in $L^2_w(\Gamma)$. For any function $f\in L^2_w(\Gamma)$ the following expansion holds

\begin{eqnarray}\nonumber
f(x)\cong \sum_{k=-N}^{+N}f_k \widehat{H}_k(x),
\end{eqnarray}
with
\begin{eqnarray}\nonumber
f_k=\frac{<f(x),\widehat{H}_k(x)>_{w(x)}}{\parallel \widehat{H}_k(x){\parallel}_{w(x)}^2}.
\end{eqnarray}
Now we can define an orthogonal projection based on the transformed Hermite functions as given below:
Let
\begin{eqnarray}\nonumber
\widehat{\mathcal{H}}_N=span\{\widehat{H}_0(x),\widehat{H}_1(x),...,\widehat{H}_n(x)\}.
\end{eqnarray}
The $L^{2}{(\Gamma)}$-orthogonal projection $\hat{\xi}_N:L^{2}{(\Gamma)}\longrightarrow\widehat{\mathcal{H}}_N$ is a mapping in a way that for any $y\in L^{2}{(\Gamma)}$,
\begin{eqnarray}\nonumber
<\hat{\xi}_N{y}-y,\phi>=0, \quad \forall \phi\in \widehat{\mathcal{H}}_N,
\end{eqnarray}
or equivalently,
\begin{eqnarray}\label{operatorHFT}
\hat{\xi}_N{y(x)}=\sum_{i=0}^{N}\hat{a}_{i}\widehat{H}_i(x).
\end{eqnarray}

%\bigskip
\subsection{Domain scaling}
It has already been mentioned in \cite{Liu.Liu.Tang} that when using a spectral approach on the whole real line $\mathbb{R}$ one can possibly increase the accuracy of the computation by a suitable scaling of the underlying time variable $t$. For example, if $y$ denotes a solution of the ordinary differential equation, then the rescaled function is
$\tilde{y}(t)=y(\frac{t}{l})$, where $l$ is constant.
Domain scaling is used in several applications presented in next section. For more details we refer the interested reader to \cite{Tang.1993}.
%/////////////////////////////////////////////////////////////////////////////
\section{Applications}\label{Applications}
In this section we apply Hermite functions collocation (HFC) method for the computation of Lane-Emden type
equations based on the transformed Hermite functions.
In general the Lane-Emden type equations are formulated as
\begin{equation}\label{GeneralLane.EmdenEQ}
y''(x)+\frac{\alpha}{x}y(x)+f(x)g(y)=h(x), \quad \alpha x\geq 0,
\end{equation}
with initial conditions
\begin{eqnarray}\nonumber
a. \quad y(0)=A,\\
b. \quad y'(0)=B,
\end{eqnarray}
where $\alpha$, $A$ and $B$ are real constants and $f(x)$, $g(y)$ and $h(x)$ are some given functions.
For other special forms of $g(y)$, the well-known Lane-Emden equations were used to model several phenomena in mathematical physics and astrophysics such as the theory of stellar structure, the thermal
behavior of a spherical cloud of gas, isothermal gas spheres and the theory of
thermionic currents \cite{Chandrasekhar,Richardson}.

In this section we apply the Hermite functions collocation method to solve some
well-known Lane-Emden type equations for various $f(x)$, $g(y)$, $A$ and $B$,
in two cases homogeneous ($h(x)=0$) and non-homogeneous ($h(x)\neq0$).

The Hermite functions are not differentiable at the point $x=0$, therefore to satisfy the second boundary condition we can multiply the operator (\ref{operatorHFT}) by $x$ and add it with $Bx$, and also for satisfying the first boundary condition we add it with $A$ as follows:
\begin{eqnarray}\nonumber
\widehat{\xi}_{N}y(x)= A+Bx+{x}\hat{\xi}_Ny(x).
\end{eqnarray}
Now for boundary conditions we have $\widehat{\xi}_{N}y(x)=A$ and $\frac{d}{dx}\widehat{\xi}_{N}y(x)=B$ when $x$ tends to zero.\\
To apply the collocation method, we construct the residual function by substituting $y(x)$ by $\widehat{\xi}_{N}y(x)$ in the Lane-Emden Eq. (\ref{GeneralLane.EmdenEQ}):
\begin{eqnarray}\nonumber
Res(x)=\frac{d^2}{dx^2}\widehat{\xi}_{N}y(x)+\frac{\alpha}{x}\frac{d}{dx}\widehat{\xi}_{N}y(x)+f(x)g(\widehat{\xi}_{N}y(x))-h(x).
\end{eqnarray}
The equations for obtaining the coefficient $a_i$s arise from equalizing $Res(x)$ to zero at $N+1$ transformed Hermite-Gauss points by Eq. (\ref{inverseTransform}), i.e, transformed roots of Hermite $H_{N+1}(x)$:
\begin{equation}\label{resforGeneralLane-emden}
Res(x_j)=0, \qquad j=0,1,2,...,N.
\end{equation}
Solving this set of equations we have the approximating function $\widehat{\xi}_{N}y(x)$. We note that these $N+1$ equations generate a set of $N+1$ nonlinear equations which can be solved by a well-known method such as the Newton method for unknown coefficients $a_i$s.
\subsection{The homogeneous Lane-Emden type equations}
%/////////////////////////////////////////////////////////////////////////
\subsubsection{Example 1 \textsl{(The standard Lane-Emden equation)}}
For $f(x) = 1$, $g(y) = y^m$, $A=1$ and $B=0$,  Eq. (\ref{GeneralLane.EmdenEQ}) is the standard Lane-Emden equation that was used to model the thermal behavior of a spherical cloud of gas acting under the mutual attraction of its molecules and subject to the classical laws of thermodynamics \cite{Shawagfeh,davis}.
\begin{equation}\label{mainLane.EmdenEQ}
y''(x)+\frac{2}{x}y'(x)+y^{m}(x)=0, \quad x\geq 0,
\end{equation}
subject to the boundary conditions
\begin{eqnarray}\nonumber
\quad y(0)=1,\\\nonumber
\quad y'(0)=0,
\end{eqnarray}
where $m\geq0$ is constant.
%%%%%%%%%%%%%%%%%%%%%%%%%%%%%%%%%%%
Substituting $m=0$, $1$ and $5$ into Eq. (\ref{mainLane.EmdenEQ}) leads to the exact solution
\begin{eqnarray}\nonumber
y(x)=1-\frac{1}{3!}x^2, \quad y(x)=\frac{sin(x)}{x} \quad\text{and} \quad y(x)=\left(1+\frac{x^2}{3}\right)^{-1/2},
\end{eqnarray}
respectively. In other cases there aren't any analytic exact solutions.
Therefore, we apply the HFC method to solve the standard Lane-Emden Eq. (\ref{mainLane.EmdenEQ}) for $m=1.5$, $2$, $2.5$, $3$ and $4$.
To this way now we can construct the residual function as follows:
\begin{eqnarray}\nonumber
Res_l(x)=\frac{d^2}{dx^2}\widehat{\xi}_{N}y(x/l)+\frac{2}{x}\frac{d}{dx}\widehat{\xi}_{N}y(x/l)+(\widehat{\xi}_{N}y(x/l))^m,
\end{eqnarray}
where $l$ is a constant that is already defined in the domain scaling description.
As said before, to obtain the coefficients $a_i$s, $Res_l(x)$ is equalized to zero at $N+1$ transformed Hermite-Gauss points by Eq. (\ref{inverseTransform}):
\begin{eqnarray}\nonumber
Res_l(x_j)=0, \qquad j=0,1,2,...,N.
\end{eqnarray}
By solving this set of equations, we can find the approximating function $\widehat{\xi}_{N}y(x)$.
%-------------------------------------------------------------------------------------

Table \ref{Tab.STLaneEmden.Zero} shows the comparison of the first zeros of the standard Lane-Emden equations, from the present method and exact values given by Horedt \cite{Horedt} for $m=1.5$, $2$, $2.5$, $3$ and $4$, respectively.

Tables \ref{Tab.STLaneEmden.values3}, \ref{Tab.STLaneEmden.values4} show the approximations of $y(x)$ for the standard Lane-Emden for $m=3,4$ respectively obtained by the method proposed in this paper and those obtained by Horedt \cite{Horedt}.

\begin{figure}
\centerline{\includegraphics[scale=0.5]{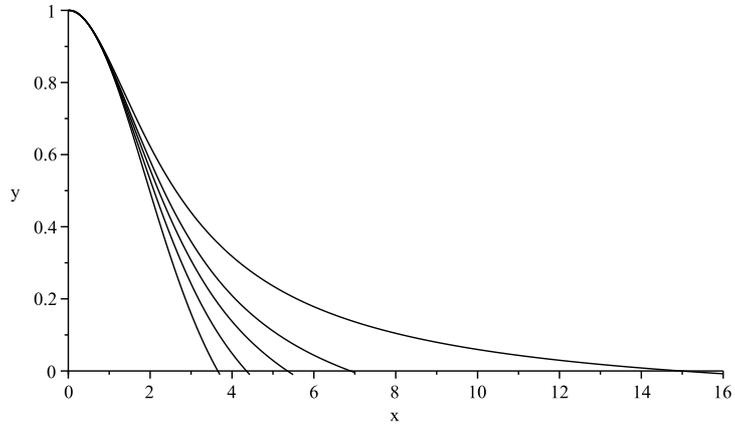}}
\caption{Graph of standard Lane-Emden equation for $m=1.5$, $2$, $2.5$, $3$ and $4$}
\label{FigAllmExpl1}
\end{figure}

\begin{figure}
\centerline{\includegraphics[scale=0.4]{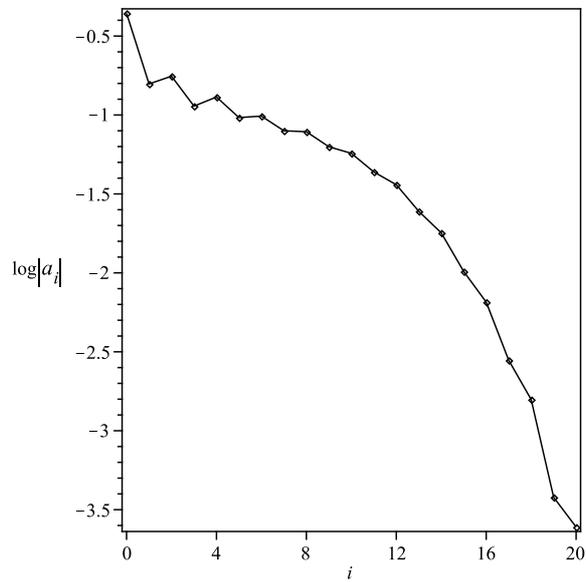}}
\caption{Logarithmic graph of absolute coefficients $|a_i|$ of Hermite function of standard Lane-Emden for $m=3$}
\label{FigCoeffExpl1m3}
\end{figure}

Table \ref{Tab.STLanemden.coef} represents the coefficients of the Hermite functions obtained by the present method
for $m=2,3$ and $4$ of the standard Lane-Emden equation.
The resulting graph of the standard Lane-Emden equation for $m=1.5$, $2$, $2.5$, $3$ and $4$ is shown in Figure \ref{FigAllmExpl1}.
The logarithmic graph of absolute coefficients of Hermite functions of standard Lane-Emden for $m=3$ is shown in Fig. \ref{FigCoeffExpl1m3} and the coefficients in Table \ref{Tab.STLanemden.coef} show that the new method has an appropriate convergence rate.
%/////////////////////////////////////////////////////////////////////////////////
\subsubsection{Example 2. \textsl{(The isothermal gas spheres equation)}}
For $f(x) = 1$, $g(y) = e^y$, $A=0$ and $B=0$,  Eq. (\ref{GeneralLane.EmdenEQ}) is the isothermal gas sphere equation.
\begin{equation}\label{EQIsothermal}
y''(x)+\frac{2}{x}y'(x)+e^{y(x)}=0, \quad x\geq 0,
\end{equation}
subject to the boundary conditions
\begin{eqnarray}\nonumber
\quad y(0)=0,\\\nonumber
\quad y'(0)=0.
\end{eqnarray}
This model can be used to view the isothermal gas spheres, where the temperature remains constant. For a thorough discussion of the formulation of Eq. (\ref{EQIsothermal}), see \cite{davis}.\\
A series solution obtained by Wazwaz \cite{Wazwaz.Appl}, Liao \cite{Liao.Appl. Math. Comput.2003}, Singh et al. \cite{Singh.Pandey2009} and Ramos \cite{Ramos.Chaos Soliton. Frac.2008} by using ADM, ADM, MHAM and series expansion respectively:
\begin{eqnarray}\label{wazwazIsothermal}
y(x)\simeq-\frac{1}{6}x^2+\frac{1}{5.4!}x^4-\frac{8}{21.6!}x^6+\frac{122}{81.8!}x^8-\frac{61.67}{495.10!}x^{10}.
\end{eqnarray}
%%However, Eq. (\ref{wazwazIsothermal}) is valid in the restricted region $0\leq x < 3.5$ \cite{Singh.Pandey2009}.
We recall that this equation has been solved by \cite{ChowdhuryHashim,Aslanov.Phys. Lett. A2008,AslanovIJCM2008,BatainehNooraniHashim} with HPM, series solutions, ADM and HAM methods respectively.
We intend to apply the HFC method to solve the isothermal gas spheres Eq. (\ref{EQIsothermal}) too.
Therefore, we construct the residual function as follows:
\begin{eqnarray}\nonumber
Res_l(x)=\frac{d^2}{dx^2}\widehat{\xi}_{N}y(x/l)+\frac{2}{x}\frac{d}{dx}\widehat{\xi}_{N}y(x/l)+e^{(\widehat{\xi}_{N}y(x/l))},
\end{eqnarray}
where $l$ is a constant that is already defined in domain scaling description.
As said before, to obtain the coefficients $a_i$s, $Res_l(x)$ is equalized to zero at $N+1$ transformed Hermite-Gauss points by Eq. (\ref{inverseTransform}):
\begin{eqnarray}\nonumber
Res_l(x_j)=0, \qquad j=0,1,2,...,N.
\end{eqnarray}
By solving the set of equations, we have the approximating function $\widehat{\xi}_{N}y(x)$.
%--------------------------------------------------------------

Tables \ref{Tab.ISOThemalgasValues} shows the comparison of $y(x)$ obtained by the method proposed in this paper with ($N=30$, $l=2$ and $k=2$) and those obtained by Wazwaz \cite{Wazwaz.Appl}.

\begin{figure}
\centerline{\includegraphics[scale=0.4]{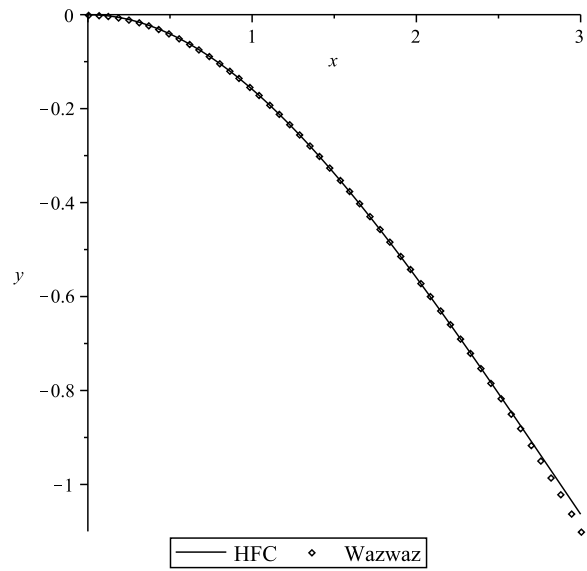}}
\caption{Graph of isothermal gas sphere equation in comparison with Wazwaz solution \cite{Wazwaz.Appl}}
\label{FiigExmpl2}
\end{figure}

\begin{figure}
\centerline{\includegraphics[scale=0.4]{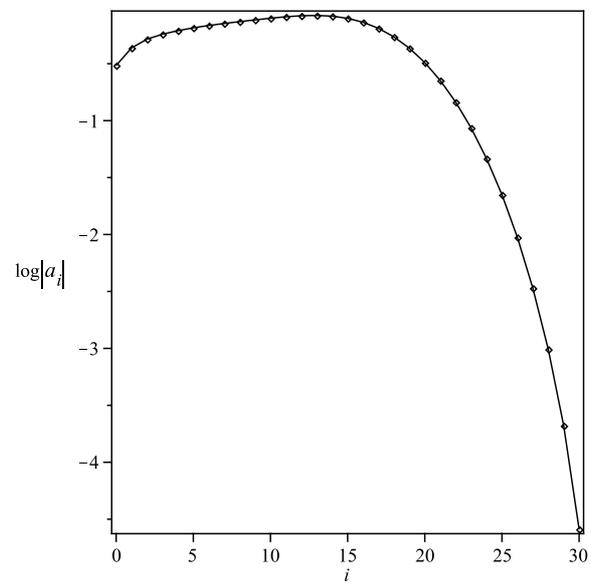}}
\caption{Logarithmic graph of absolute coefficients $|a_i|$ of Hermite function of isothermal gas sphere equation}
\label{FigCoeffExpl2}
\end{figure}

The resulting graph of the isothermal gas spheres equation in comparison to the presented method and those obtained by Wazwaz \cite{Wazwaz.Appl} are shown in Figure \ref{FiigExmpl2}.
%%%%%%%
The logarithmic graph of the absolute coefficients of Hermite functions of the standard isothermal gas spheres is shown in Figure \ref{FigCoeffExpl2}. This graph shows that the new method has an appropriate convergence rate.
%//////////////////////////////////////////////////////////////////////////////
\subsubsection{Example 3.}
For $f(x) = 1$, $g(y) = \sinh(y)$, $A=1$ and $B=0$, Eq. (\ref{GeneralLane.EmdenEQ}) will be one of the Lane-Emden type equations that is we want to solve:

\begin{equation}\label{EQSinh.y}
y''(x)+\frac{2}{x}y'(x)+\sinh(y)=0, \quad x\geq 0,
\end{equation}
subject to the boundary conditions
\begin{eqnarray}\nonumber
\quad y(0)=1,\\\nonumber
\quad y'(0)=0.
\end{eqnarray}
A series solution obtained by Wazwaz~\cite{Wazwaz.Appl} by using Adomian Decomposition Method (ADM) is:
\begin{align}\nonumber
y(x)\simeq & 1-\,{\frac { \left( {e^{2}}-1 \right) {x}^{2}}{12{e}}}+{\frac {1
}{480}}\,{\frac { \left( {e^{4}}-1 \right) {x}^{4}}{{e^{2}}}}-{\frac {
1}{30240}}\,{\frac { \left( 2\,{e^{6}}+3\,{e^{2}}-3\,{e^{4}}-2
 \right) {x}^{6}}{{e^{3}}}}\\\nonumber
&+{\frac {1}{26127360}}\,{\frac { \left( 61
\,{e^{8}}-104\,{e^{6}}+104\,{e^{2}}-61 \right) {x}^{8}}{{e^{4}}}}.
\end{align}

We intend to apply HFC method to solve equation Eq. (\ref{EQSinh.y}).\\
Therefore, we construct the residual function as follows:
\begin{eqnarray}\nonumber
Res_l(x)=\frac{d^2}{dx^2}\widehat{\xi}_{N}y(x/l)+\frac{2}{x}
\frac{d}{dx}\widehat{\xi}_{N}y(x/l)+\sinh{(\widehat{\xi}_{N}y(x/l))},
\end{eqnarray}
where $l$ is a constant that is already defined in domain scaling description.
To obtain the coefficients $a_i$s, $Res_l(x)$ is equalized to zero at $N+1$ transformed Hermite-Gauss points by Eq. (\ref{inverseTransform}):
\begin{eqnarray}\nonumber
Res_l(x_j)=0, \qquad j=0,1,2,...,N.
\end{eqnarray}
By solving this set of equations, we have the approximating function $\widehat{\xi}_{N}y(x)$.
%--------------------------------------------------------------

\begin{figure}
\centerline{\includegraphics[scale=0.4]{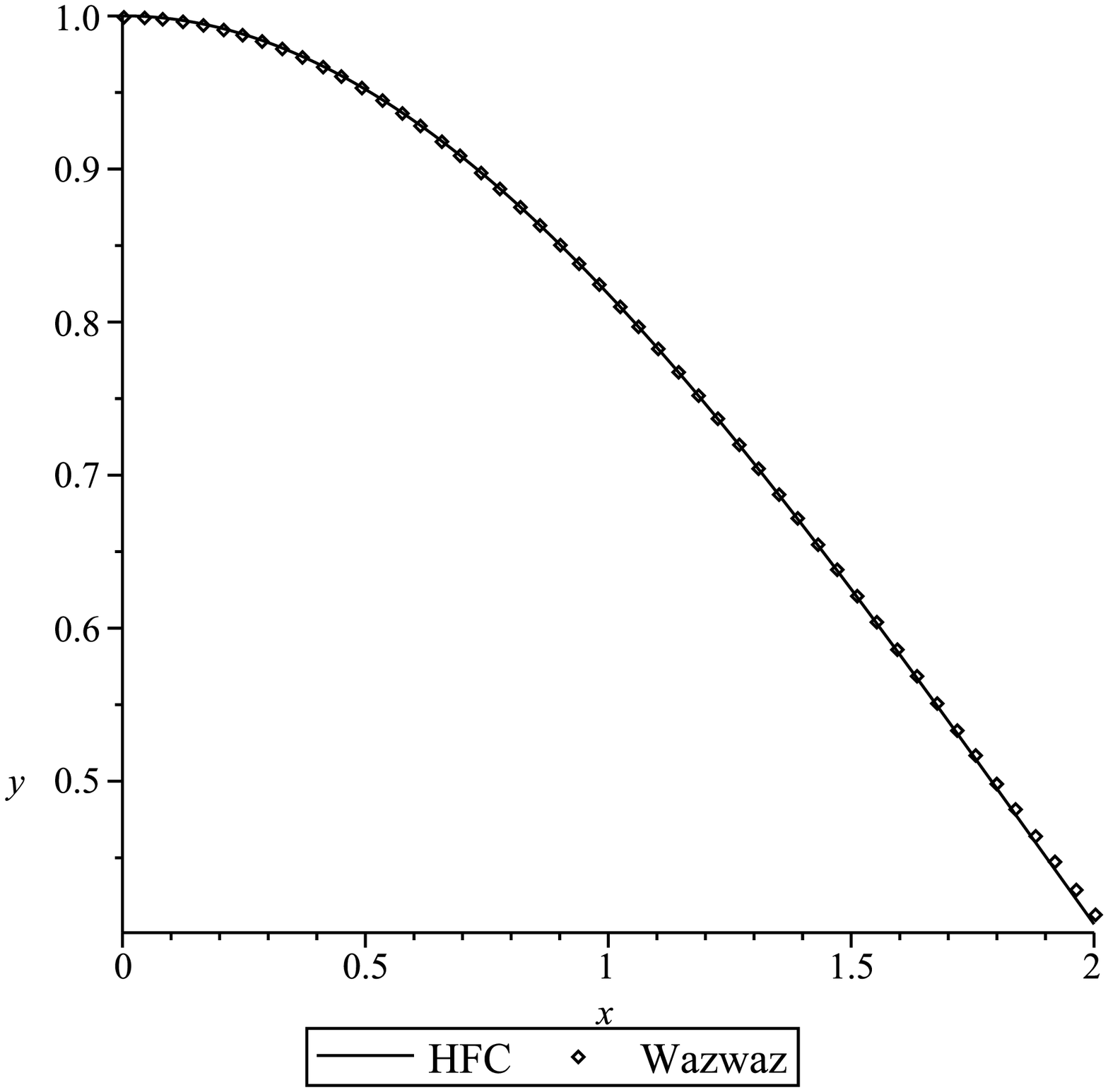}}
\caption{Graph of equation example 3 in comparing the presented method and Wazwaz solution \cite{Wazwaz.Appl}}
\label{FigExmpl3}
\end{figure}

Tables \ref{Tab.Sinh.y.Values} shows the comparison of $y(x)$ obtained by the new method proposed in this paper with ($n=10$, $k=1$ and $l=2$), and those obtained by Wazwaz \cite{Wazwaz.Appl}.
The resulting graph of Eq. (\ref{EQSinh.y}) in comparison to the presented method and those obtained by Wazwaz \cite{Wazwaz.Appl} are shown in Figure \ref{FigExmpl3}.
%////////////////////////////////////////////////////////////////////////////////
\subsubsection{Example 4.}
For $f(x) = 1$, $g(y) = \sin(y)$, $A=1$ and $B=0$, Eq. (\ref{GeneralLane.EmdenEQ}) will be one of the Lane-Emden type equations that is we would like to solve:
\begin{equation}\label{EQSin.y}
y''(x)+\frac{2}{x}y'(x)+\sin(y)=0, \quad x\geq 0,
\end{equation}
subject to the boundary conditions
\begin{eqnarray}\nonumber
\quad y(0)=1,\\\nonumber
\quad y'(0)=0.
\end{eqnarray}
A series solution obtained by Wazwaz \cite{Wazwaz.Appl} by using ADM is:
\begin{align}\nonumber
y(x)\simeq & 1-\frac{1}{6}k_1x^2+\frac{1}{120}k_1k_2x^4+k_1(\frac{1}{3024}k_1^2-\frac{1}{5040}k_2^2)x^6\\\nonumber
&+k_1k_2(-\frac{113}{3265920}k_1^2+\frac{1}{362880}k_2^2)x^8\\\nonumber
&+k_1(\frac{1781}{898128000}k_1^2k_2^2-\frac{1}{399168000}k_2^4-\frac{19}{23950080}k_1^4)x^{10},
\end{align}
where $k_1=\sin(1)$ and $k_2=\cos(1)$.\\

We intend to apply the HFC method to solve this type equation (Eq. \ref{EQSin.y}).\\
Therefore, we construct the residual function as follows:
\begin{eqnarray}\nonumber
Res_l(x)=\frac{d^2}{dx^2}\widehat{\xi}_{N}y(x/l)+\frac{2}{x}\frac{d}{dx}\widehat{\xi}_{N}y(x/l)+\sin{(\widehat{\xi}_{N}y(x/l))},
\end{eqnarray}
where $l$ is a constant that is already defined in domain scaling description.
As said before, to obtain the coefficients $a_i$s, $Res_l(x)$ is equalized to zero at $N+1$ transformed Hermite-Gauss points by Eq. (\ref{inverseTransform}):
\begin{eqnarray}\nonumber
Res_l(x_j)=0, \qquad j=0,1,2,...,N.
\end{eqnarray}
By solving this set of equations, we have the approximating function $\widehat{\xi}_{N}y(x)$.
%--------------------------------------------------------------

\begin{figure}
\centerline{\includegraphics[scale=0.4]{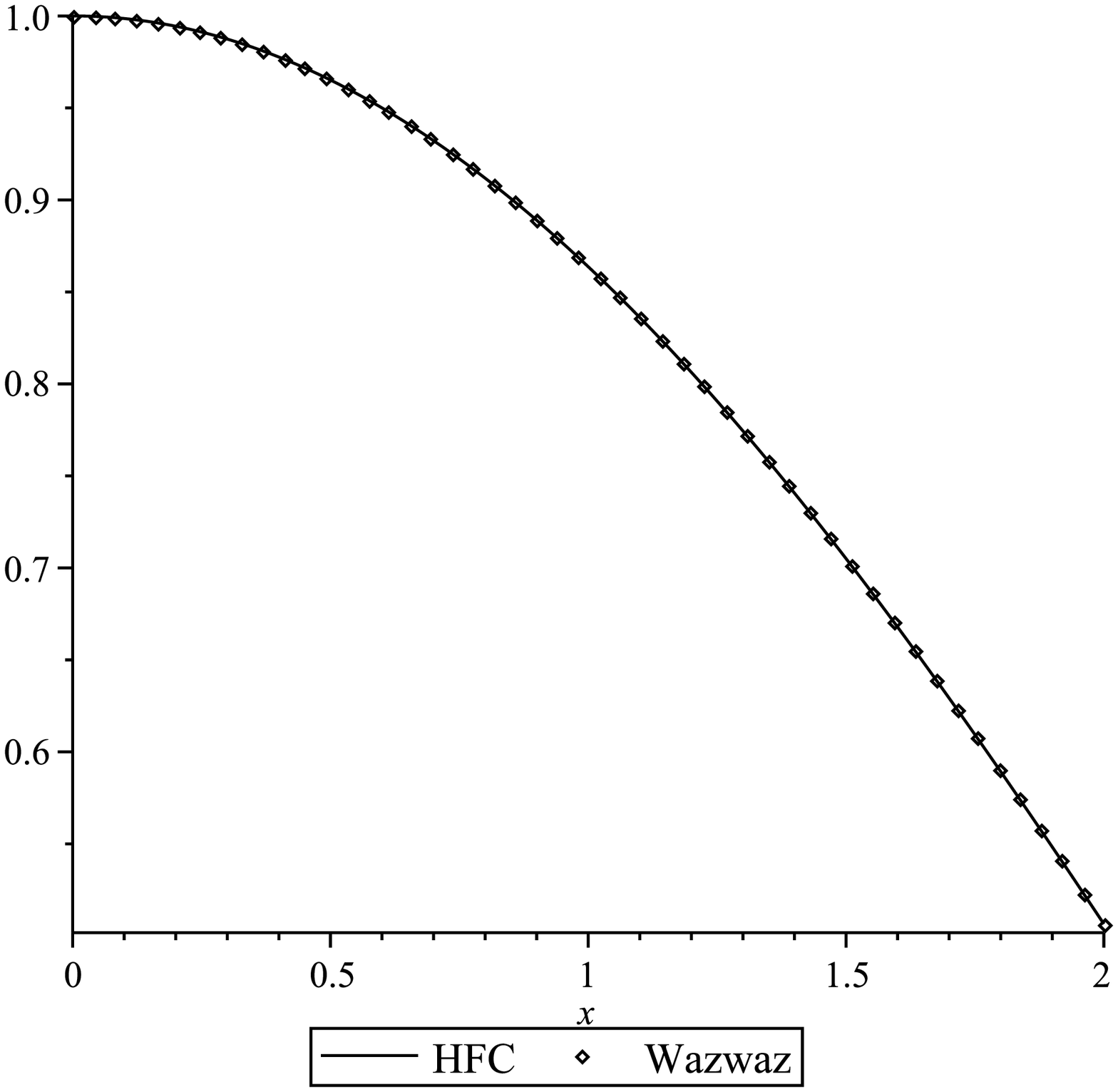}}
\caption{Graph of equation example 4 in comparing the presented method and Wazwaz solution \cite{Wazwaz.Appl}}
\label{FigExmpl5}
\end{figure}

Tables \ref{Tab.Sin.y.Values} shows the comparison of $y(x)$ obtained by the method proposed in this paper with ($n=15$, $k=1$ and $l=2$), and those obtained by Wazwaz \cite{Wazwaz.Appl}.
In order to compare the present method with those obtained by Wazwaz~\cite{Wazwaz.Appl}) the resulting
graph of  Eq. (\ref{EQSin.y}) is shown in Figure~\ref{FigExmpl5}.

%////////////////////////////////////////////////////////////////////////////////////////
\subsubsection{Example 5.}
For $f(x) = 1$, $g(y) = 4(2e^y+e^{y/2})$, $A=0$ and $B=0$, Eq. (\ref{GeneralLane.EmdenEQ}) will be one of the Lane-Emden type equations that is to solve.
\begin{equation}\label{EQee.y}
y''(x)+\frac{2}{x}y'(x)+4(2e^y+e^{y/2})=0, \quad x\geq 0,
\end{equation}
subject to the boundary conditions
\begin{eqnarray}\nonumber
\quad y(0)=0,\\\nonumber
\quad y'(0)=0.
\end{eqnarray}
which has the following analytical solution:
\begin{equation}\label{EQee.exact}
y(x)=-2ln(1+x^2).
\end{equation}
This type of equation has been solved by \cite{Yildirim,ChowdhuryHashimPhysLetA} with VIM and HPM methods respectively.

We applied the HFC method to solve equation Eq. (\ref{EQee.y}).
Therefore, we construct the residual function as follows:
\begin{eqnarray}\nonumber
Res_l(x)=\frac{d^2}{dx^2}\widehat{\xi}_{N}y(x/l)+\frac{2}{x}\frac{d}{dx}\widehat{\xi}_{N}y(x/l)+4(2e^{\widehat{\xi}_{N}y(x/l)}+e^{{\widehat{\xi}_{N}y(x/l)}/2}),
\end{eqnarray}
where $l$ is a constant that is already defined in domain scaling description.
To obtain the coefficients $a_i$s, $Res_l(x)$ is equalized to zero at $N+1$ Hermite-Gauss points, i.e, roots of Hermite $H_{N+1}(x)$:
\begin{equation}\nonumber
Res_l(x_j)=0, \qquad j=0,1,2,...,N.
\end{equation}
By solving the set of equations, we have the approximating function $\widehat{\xi}_{N}y(x)$.
%-------------------------------------------------------------------------------

\begin{figure}
\centerline{\includegraphics[scale=0.4]{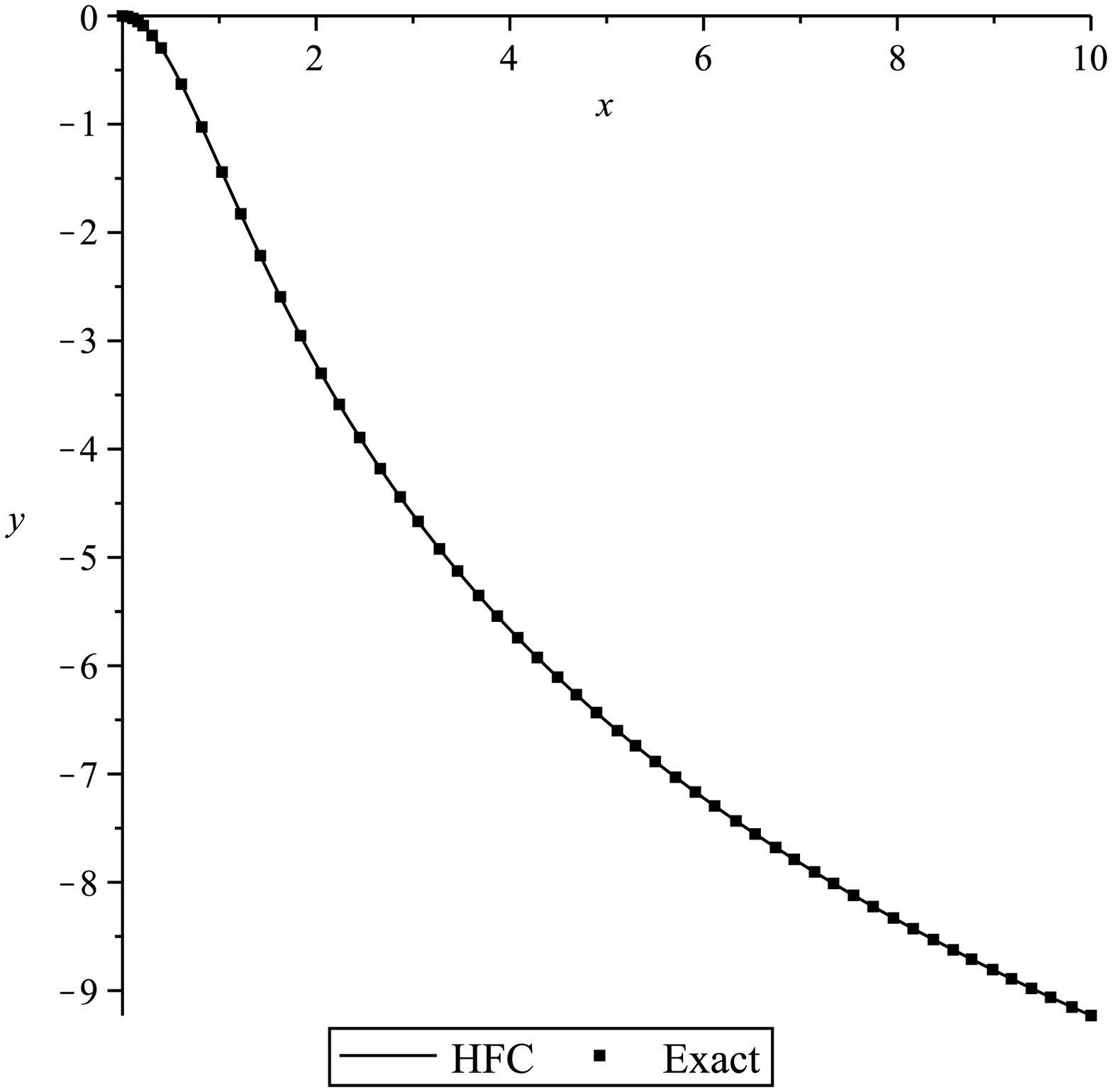}}
\caption{Graph of equation example 5 in comparing the presented method and analytic solution}
\label{FigExmpl6}
\end{figure}

Tables \ref{Tab.ee.y.Values} shows the comparison of $y(x)$ obtained by the new method proposed in this paper with ($n=30$, $k=2/3$ and $l=2$) and the analytic solution Eq. (\ref{EQee.exact}).
In order to compare the present method with the analytic solution, the resulting graph of Eq. (\ref{EQee.y})
is shown in Figure~\ref{FigExmpl6}.

%////////////////////////////////////////////////////////////////////////////////////////
\subsubsection{Example 6.}
For $f(x) = 1$, $g(y) = -6y-4y\ln(y)$, $A=1$ and $B=0$, Eq. (\ref{GeneralLane.EmdenEQ}) will be one of the Lane-Emden type equations that is:
\begin{equation}\label{EQ4yln.y}
y''(x)+\frac{2}{x}y'(x)-6y(x)=4y(x)\ln(y(x)), \quad x\geq 0,
\end{equation}
subject to the boundary conditions
\begin{eqnarray}\nonumber
\quad y(0)=1,\\\nonumber
\quad y'(0)=0,
\end{eqnarray}
which has the following analytical solution:
\begin{equation}\label{EQ4yln.exact}
y(x)=e^{x^2}.
\end{equation}
This type of equation has been solved by \cite{Yildirim,Ramos.Appl. Math. Comput.2005} with VIM and linearization methods respectively.
Now we apply the HFC method to solve this type of equation but in this model we have $y(x)\ln(y(x))$ term that increases
the order of calculation; therefore, we can use the transform $y(x)=e^{z(x)}$ in which $z(x)$ is unknown; where
upon transformed form of the model will become as follows:
\begin{equation}\label{TransEQ4yln.y}
z''(x)+(z'(x))^2+\frac{2}{x}z'(x)-6=4z(x), \quad x\geq 0,
\end{equation}
with the boundary conditions
\begin{eqnarray}\nonumber
\quad z(0)=0,\\\nonumber
\quad z'(0)=0.
\end{eqnarray}
We applied the HFC method to solve this type of equation (i. e. Eq. (\ref{TransEQ4yln.y})).\\
Therefore, we construct the residual function as follows:
\begin{eqnarray}\nonumber
Res_l(x)=\frac{d^2}{dx^2}\widehat{\xi}_{N}z(x/l)+\left(\frac{d}{dx}\widehat{\xi}_{N}z(x/l)\right)^2
+\frac{2}{x}\frac{d}{dx}\widehat{\xi}_{N}z(x/l)-4(\widehat{\xi}_{N}z(x/l))-6,
\end{eqnarray}
where $l$ is a constant that is defined in domain scaling description, before.
To obtain the coefficients $a_i$s, $Res_l(x)$ is equalized to zero at $N+1$ transformed Hermite-Gauss
points by Eq. (\ref{inverseTransform}):
\begin{eqnarray}\nonumber
Res_l(x_j)=0, \qquad j=0,1,2,...,N.
\end{eqnarray}
By solving this set of equations, we have the approximating function $\widehat{\xi}_{N}z(x)$ and also $\widehat{\xi}_{N}y(x)$.
%--------------------------------------------------------------

\begin{figure}
\centerline{\includegraphics[scale=0.4]{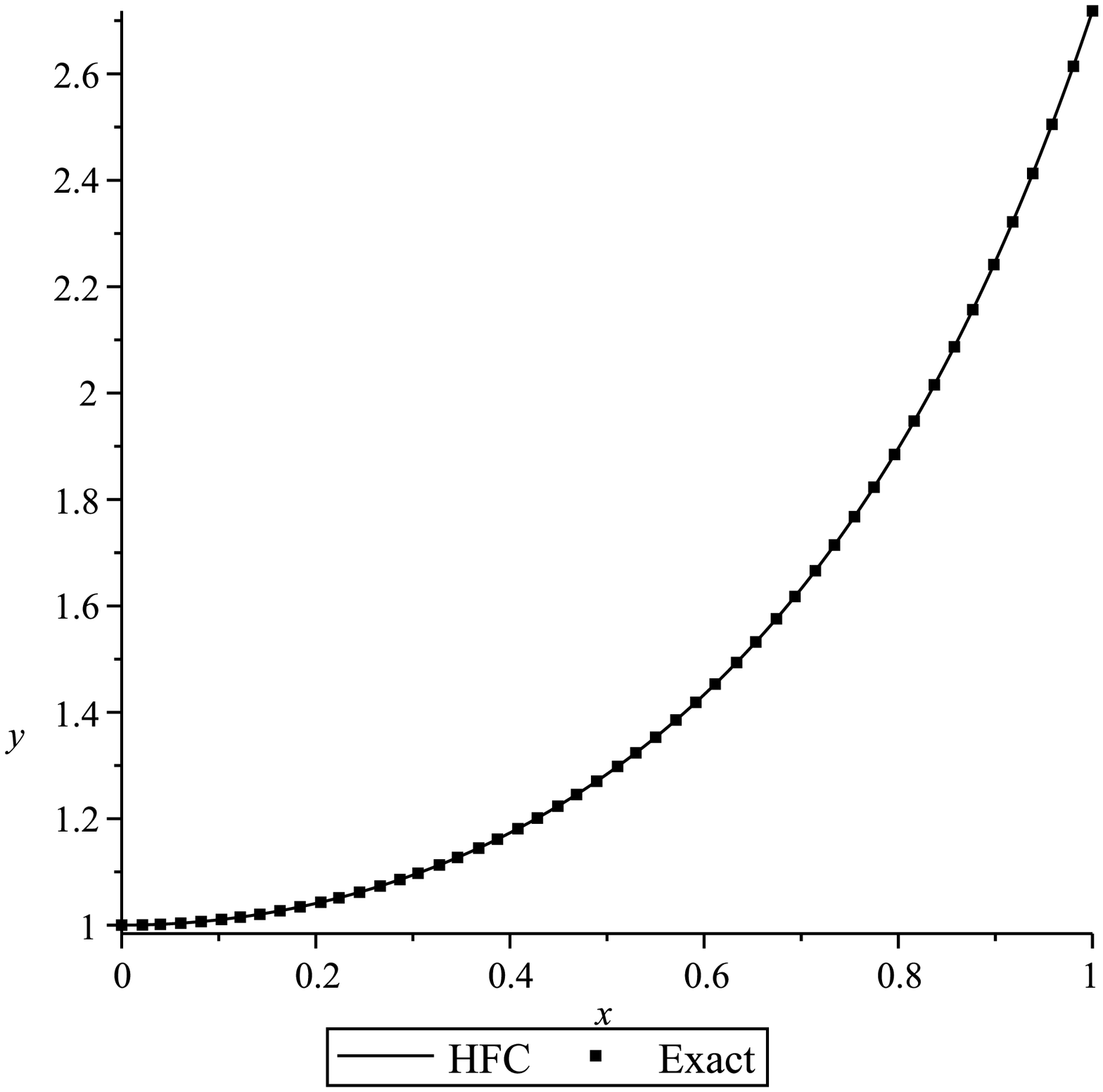}}
\caption{Graph of equation example 6 in comparing the presented method and analytic solution}
\label{FigExmpl7}
\end{figure}

Tables \ref{Tab.4yln.y.Values} shows the comparison of $y(x)$ obtained by the method proposed in this paper with ($n=30$, $k=6$ and $L=2$), and the analytic solution i. e. Eq. (\ref{EQ4yln.exact}).
The resulting graph of the Eq. (\ref{EQ4yln.y}) with presented method is shown in Figure \ref{FigExmpl7}
to make it easier to compare with the analytic solution .

\begin{figure}
\centerline{\includegraphics[scale=0.4]{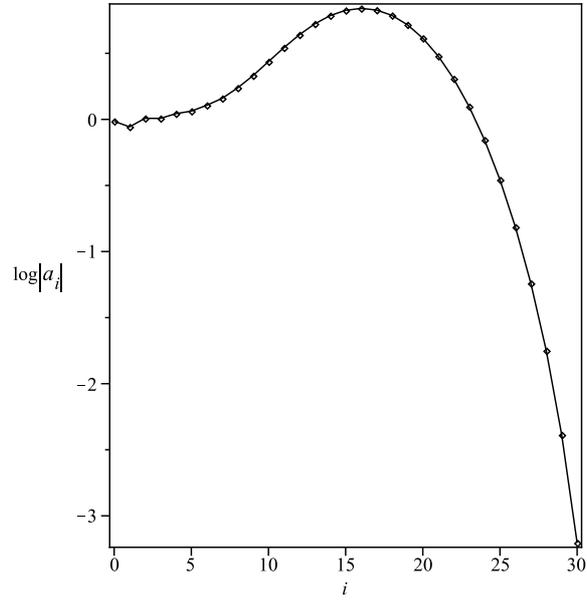}}
\caption{Logarithmic graph of absolute coefficients $|a_i|$ of Hermite function of example 6}
\label{FigCoeffExpl7}
\end{figure}

The logarithmic graph of absolute coefficients of Hermite functions of the standard isothermal gas spheres is shown in Figure \ref{FigCoeffExpl7}. This graph shows that the new method has an appropriate convergence rate.
%////////////////////////////////////////////////////////////////////////////////////////
\subsubsection{Example 7.}
For $f(x) = -2(2x^2+3)$, $g(y) = y$, $A=1$ and $B=0$, Eq. (\ref{GeneralLane.EmdenEQ}) will be one of the Lane-Emden type equations that is
\begin{equation}\label{EQex2.y}
y''(x)+\frac{2}{x}y'(x)-2(2x^2+3)y=0, \quad x\geq 0,
\end{equation}
subject to the boundary conditions
\begin{eqnarray}\nonumber
\quad y(0)=1,\\\nonumber
\quad y'(0)=0,
\end{eqnarray}
which has the following analytical solution:
\begin{equation}\label{EQex2.exact}
y(x)=e^{x^2}.
\end{equation}
This type of equation has been solved by \cite{Ramos.Appl. Math. Comput.2005,Yildirim,ChowdhuryHashimPhysLetA} with linearization, VIM and HPM methods respectively.
%%%%%%
We applied the HFC method to solve the equation ( \ref{EQex2.y}).\\
Therefore, we construct the residual function as follows:
\begin{eqnarray}\nonumber
Res_l(x)=\frac{d^2}{dx^2}\widehat{\xi}_{N}y(x/l)+\frac{2}{x}\frac{d}{dx}\widehat{\xi}_{N}y(x/l)-2(2x^2+3)\widehat{\xi}_{N}y(x/l),
\end{eqnarray}
where $l$ is a constant that is already defined in domain scaling description.
As said before, to obtain the coefficients $a_i$s, $Res_l(x)$ is equalized to zero at $N+1$ transformed Hermite-Gauss points by Eq. (\ref{inverseTransform}):
\begin{eqnarray}\nonumber
Res_l(x_j)=0, \qquad j=0,1,2,...,N.
\end{eqnarray}
By solving this set of equations, we have the approximating function $\widehat{\xi}_{N}y(x)$.
%--------------------------------------------------------------

Tables \ref{Tab.ex2.y.Values} shows the comparison of $y(x)$ obtained by the new method proposed in this paper with ($n=30$, $k=6$ and $l=2$), and analytic solution Eq. (\ref{EQex2.exact}).

\begin{figure}
\centerline{\includegraphics[scale=0.4]{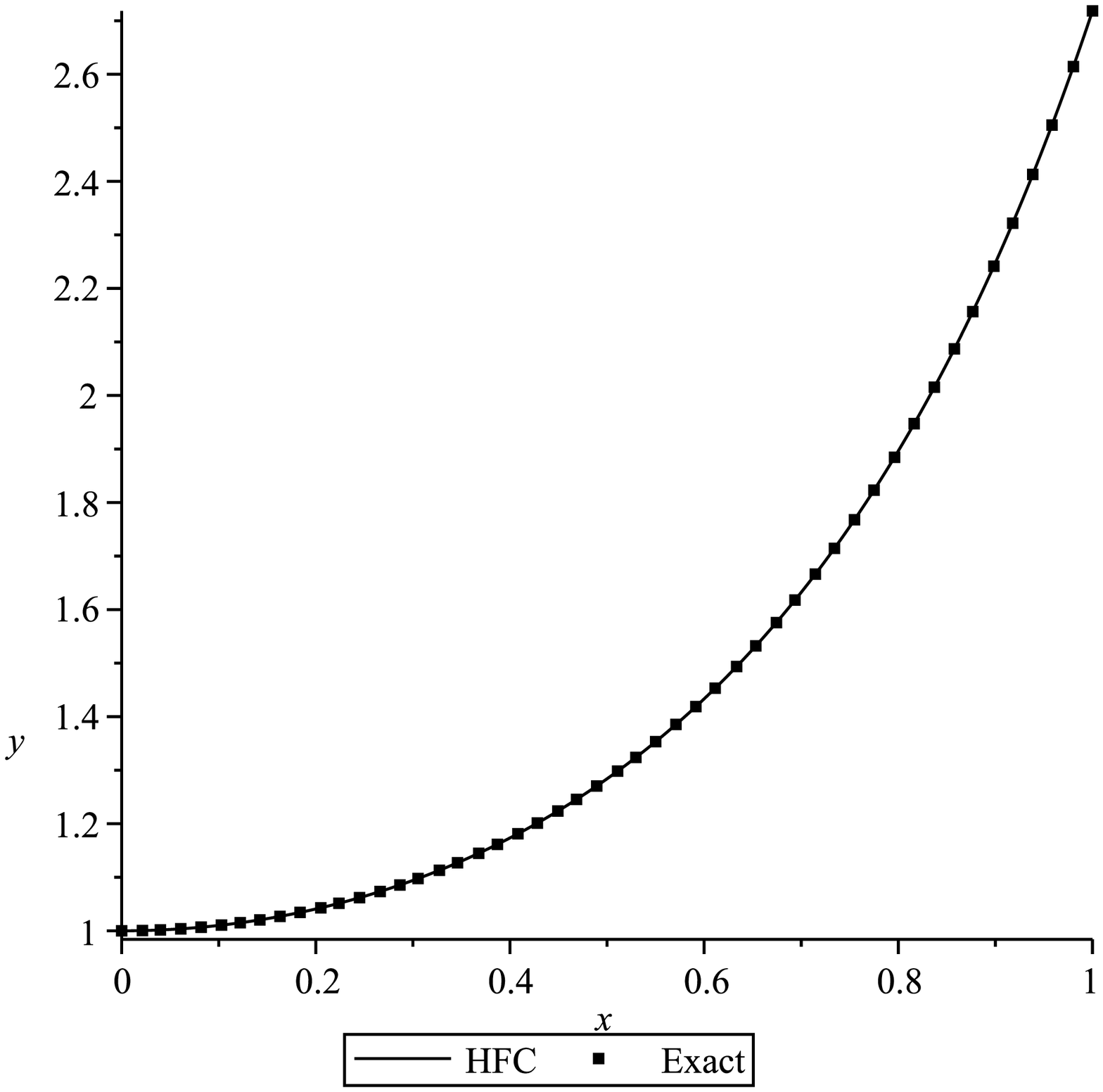}}
\caption{Graph of equation example 7 in comparing the presented method and analytic solution}
\label{FigExmpl8}
\end{figure}

\begin{figure}
\centerline{\includegraphics[scale=0.4]{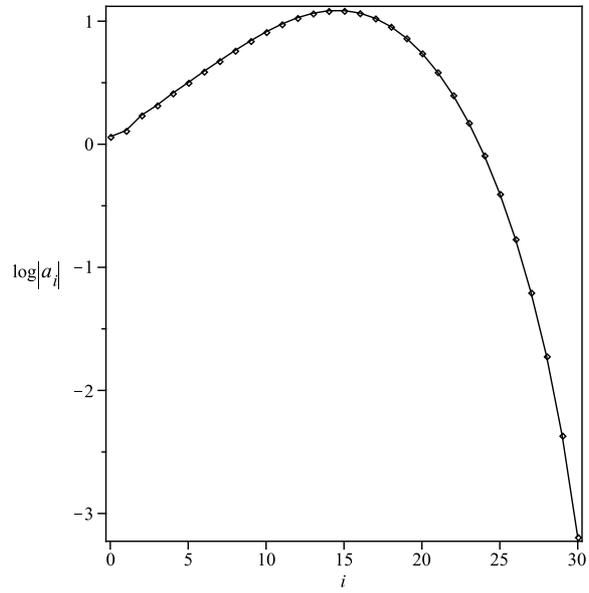}}
\caption{Logarithmic graph of absolute coefficients $|a_i|$ of Hermite function of equation of example 7}
\label{FigCoeffExpl8}
\end{figure}

The resulting graph of Eq. (\ref{EQex2.y})
is shown in Figure \ref{FigExmpl8}. Thus it will be easy for the reader to compare the new result
with the exact solution.
%%%%%%%%%%%%%%%%%%%%%%%%%%%%%
The logarithmic graph of absolute coefficients of Hermite functions of Eq. (\ref{EQex2.y}) is shown in Figure \ref{FigCoeffExpl8}. This graph shows that the current method has an appropriate convergence rate.
%--------------------------------------------------------------
\subsection{The non-homogeneous Lane-Emden type equations}
\subsubsection{Example 8.}
For $f(x) = x$, $g(y) = y$, $h(x)=x^5-x^4+44x^2-30x$, $\alpha=8$, $A=0$ and $B=0$, Eq. (\ref{GeneralLane.EmdenEQ}) will be one of the Lane-Emden type equations that is absorbing to solve.
\begin{equation}\label{EQnh1.y}
y''(x)+\frac{8}{x}y'(x)+xy(x)=x^5-x^4+44x^2-30x, \quad x\geq 0,
\end{equation}
subject to the boundary conditions
\begin{eqnarray}\nonumber
\quad y(0)=0,\\\nonumber
\quad y'(0)=0,
\end{eqnarray}
which has the following analytical solution:
\begin{equation}\label{EQnh1.exact}
y(x)=x^4-x^3.
\end{equation}
This type of equation has been solved by \cite{Ramos.Appl. Math. Comput.2005,ChowdhuryHashim,BatainehNooraniHashim,ZhangWuLuo} with linearization, HPM, HAM and TSADM methods respectively.
%%%%%%%%%%%%%%%%%%%%%%
We applied the HFC method to solve this type equation Eq. (\ref{EQnh1.y}).
Therefore, we construct the residual function as follows:
\begin{eqnarray}\nonumber
Res_l(x)=\frac{d^2}{dx^2}\widehat{\xi}_{N}y(x/l)+\frac{8}{x}\frac{d}{dx}\widehat{\xi}_{N}y(x/l)+x\widehat{\xi}_{N}y(x/l)-x^5+x^4-44x^2+30x,
\end{eqnarray}
where $l$ is a constant that is already defined in domain scaling description.
To obtain the coefficients $a_i$s, $Res_l(x)$ is equalized to zero at $N+1$ transformed Hermite-Gauss points by Eq. (\ref{inverseTransform}):
\begin{eqnarray}\nonumber
Res_l(x_j)=0, \qquad j=0,1,2,...,N.
\end{eqnarray}
By solving this set of equations, we have the approximating function $\widehat{\xi}_{N}y(x)$.
%--------------------------------------------------------------

\begin{figure}
\centerline{\includegraphics[scale=0.4]{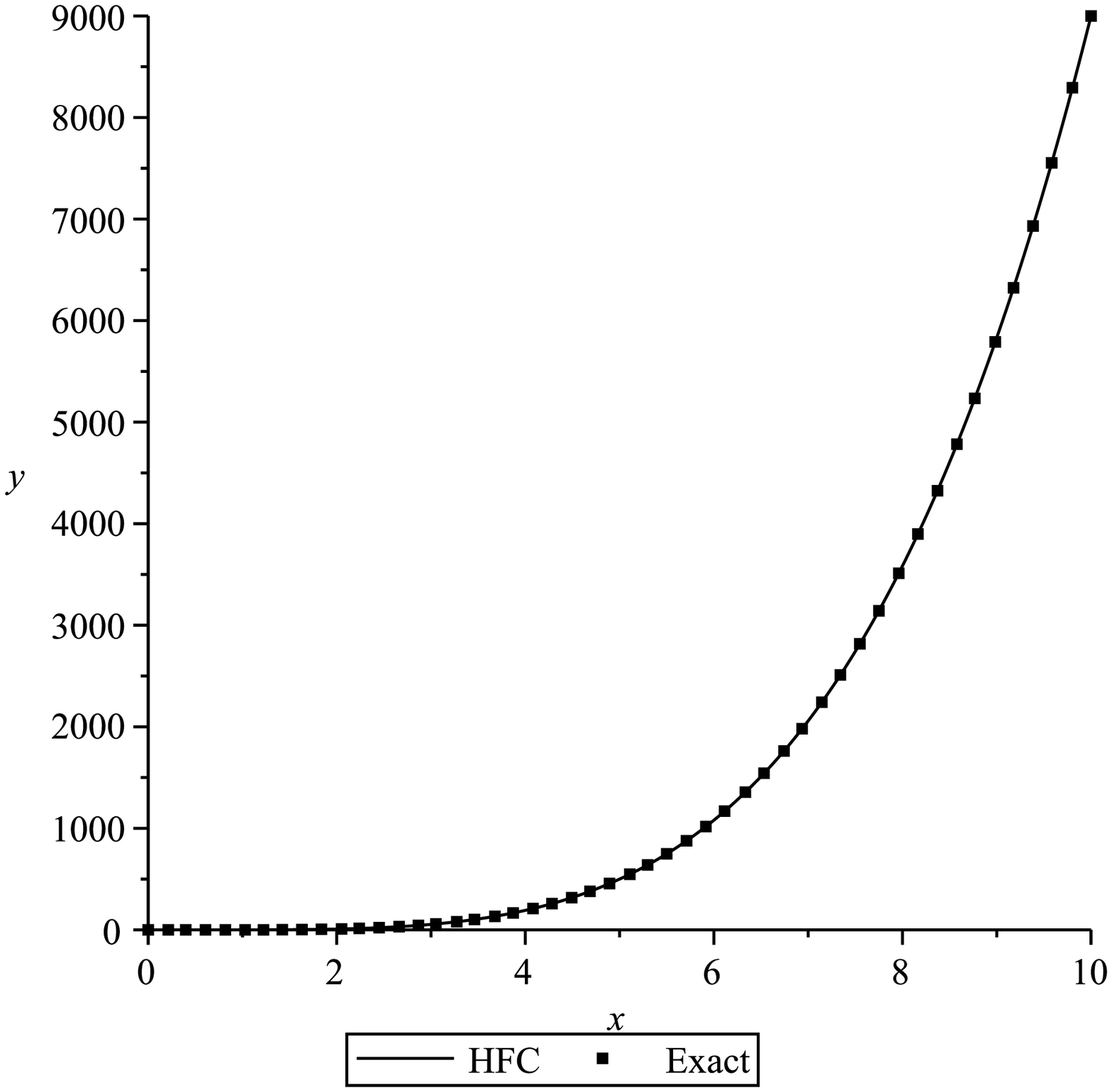}}
\caption{Graph of equation example 8 in comparing the presented method and analytic solution}
\label{FigExmpl9}
\end{figure}

Tables \ref{Tab.nh1.y.Values} shows the comparison of $y(x)$ obtained by the method proposed in this paper with ($n=30$, $k=2/3$ and $l=2$), and analytic solution Eq. (\ref{EQnh1.exact}).
%%%%%%%%%%%%%%%%%
The resulting graph of the Eq. (\ref{EQnh1.y}) for the presented method and the analytic solution are shown in Figure \ref{FigExmpl9}.
%////////////////////////////////////////////////////////////////////////////////////////
\subsubsection{Example 9.}
\begin{figure}
\centerline{\includegraphics[scale=0.4]{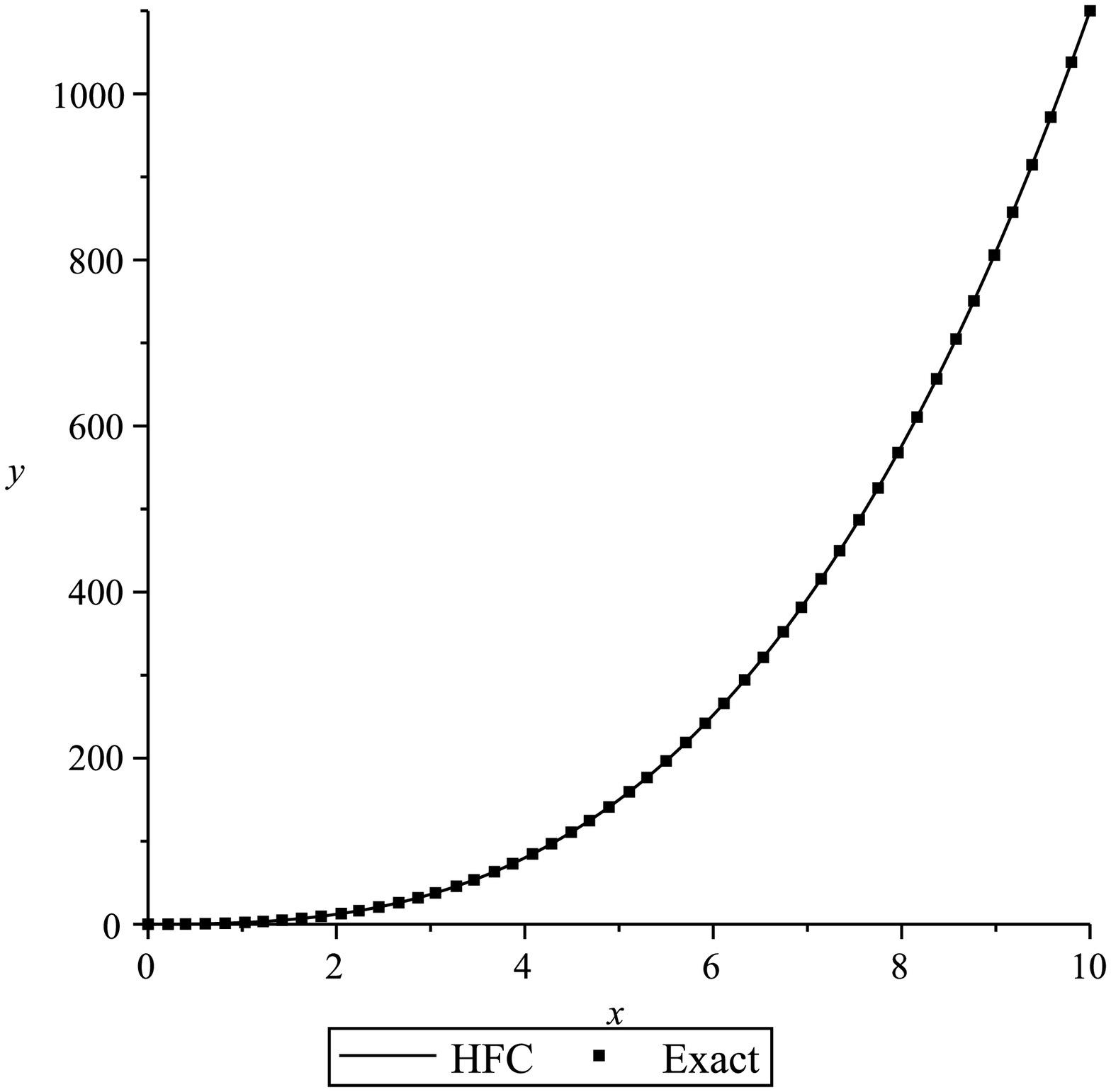}}
\caption{Graph of equation example 9 in comparing the presented method and analytic solutio}
\label{FigExmpl10}
\end{figure}

For $f(x) = 1$, $g(y) = y$, $h(x)=6+12x+x^2+x^3$, $A=0$ and $B=0$, Eq. (\ref{GeneralLane.EmdenEQ}) will be one of the Lane-Emden type equations that is
% absorbing to solve.
\begin{equation}\label{EQnh2.y}
y''(x)+\frac{2}{x}y'(x)+y(x)=6+12x+x^2+x^3, \quad x\geq 0,
\end{equation}
subject to the boundary conditions
\begin{eqnarray}\nonumber
\quad y(0)=0,\\\nonumber
\quad y'(0)=0,
\end{eqnarray}
which has the following analytical solution:
\begin{equation}\label{EQnh2.exact}
y(x)=x^2+x^3.
\end{equation}
This equation has been solved by \cite{Ramos.Appl. Math. Comput.2005,Yildirim, ChowdhuryHashimPhysLetA,ZhangWuLuo} with with linearization, VIM, HPM and TSADM methods respectively..
%%%%%%%%%%%%%%%
We applied the HFC method to solve Eq. (\ref{EQnh2.y}).\\
Therefore, we construct the residual function as follows:
\begin{eqnarray}\nonumber
Res_l(x)=\frac{d^2}{dx^2}\widehat{\xi}_{N}y(x/l)+\frac{2}{x}\frac{d}{dx}\widehat{\xi}_{N}y(x/l)+\widehat{\xi}_{N}y(x/l)-6-12x-x^2-x^3,
\end{eqnarray}
where $l$ is a constant that is already defined in domain scaling description.
As said before, to obtain the coefficients $a_i$s, $Res_l(x)$ is equalized to zero at $N+1$ transformed Hermite-Gauss points by Eq. (\ref{inverseTransform}):
\begin{eqnarray}\nonumber
Res_l(x_j)=0, \qquad j=0,1,2,...,N.
\end{eqnarray}
By solving this set of equations, we have the approximating function $\widehat{\xi}_{N}y(x)$.
%--------------------------------------------------------------
%%%%%%%%%%%%%%%
Tables \ref{Tab.nh2.y.Values} shows the comparison of $y(x)$ obtained by the method proposed in this paper with ($n=30$, $k=2/3$ and $l=2$), and analytic solution Eq. (\ref{EQnh2.exact}).
%%%%%%%%%%%%%%%%%
The resulting graph of Eq. (\ref{EQnh2.y}) with presented method in comparison to the analytic solution is shown in Figure \ref{FigExmpl10}.
%////////////////////////////////////////////////////////////////////////////////

\section{Conclusions}\label{discussion}
The Lane-Emden equations describe a variety of phenomena in theoretical physics and astrophysics, including the aspects of stellar structure, the thermal history of a spherical cloud of gas, isothermal gas spheres, and thermionic currents \cite{Chandrasekhar}.
Lane-Emden equations have been considered by many mathematicians as mentioned before~\cite{Dehghan.New Astron.2008}.
The fundamental goal of this paper has been to construct an approximation to the solution of nonlinear Lane-Emden type equations in
a semi-infinite interval. A set of Hermite functions is proposed to provide an effective
but simple way to improve the convergence of the solution by the collocation method.
The validity of the method is based on the assumption that it converges by increasing the number of collocation points.
A comparison is made among the exact solution and the numerical solutions of Horedt \cite{Horedt} and the series solutions of Wazwaz \cite{Wazwaz.Appl}, Liao \cite{Liao.Appl. Math. Comput.2003}, Singh et al. \cite{Singh.Pandey2009} and Ramos \cite{Ramos.Chaos Soliton. Frac.2008} and the current work. It has been shown that the present work provides acceptable approach for Lane-Emden type equations. Also it was confirmed
%%by the theorem and
by logarithmic figures of absolute coefficients, this approach has exponentially convergence rate.
%%%
In total an important concern of spectral methods is the choice of basis functions; the basis functions have three different properties: easy to compute, rapid convergence and completeness, which means that any solution can be represented
to arbitrarily high accuracy by taking the truncation $N$ to be sufficiently large.

\section*{Acknowledgments}
The research of first author (K. Parand) was supported by a grant from Shahid Beheshti University.
%*************************************************************************************

%////////////////////////////////////////////////////
\clearpage
\begin{table}
\caption{Comparison of the first zeros of standard Lane-Emden equations, for the present method and exact numerical values given by Horedt \cite{Horedt}}
\begin{tabular*}{\columnwidth}{@{\extracolsep{\fill}}*{6}{c}}
\hline
$m$ & $N$ & $k$ & $l$ & Present method &  Exact value\\
\hline
$1.5$ & $4$  & $1$ &  $3.74224350$ & $3.65375374$ & $3.65375374$\\
$2$   & $10$ & $1$ &  $1.97027600$ & $4.35287460$ & $4.35287460$\\
$2.5$ & $10$ & $1$ &  $1.97668316$ & $5.35527546$ & $5.35527546$\\
$3$   & $20$ & $1$ &  $1.86927585$ & $6.89684862$ & $6.89684862$\\
$4$   & $12$ & $1/3$& $1.97137830$ & $14.9715463$ & $14.9715463$\\
\hline
\end{tabular*}\label{Tab.STLaneEmden.Zero}
\end{table}
%-----------------------------------------------------------------------
\begin{table}
\caption{Comparison of $y(x)$ values of standard Lane-Emden equation, for the present method and exact values given by Horedt \cite{Horedt}, for m=3}
\begin{tabular*}{\columnwidth}{@{\extracolsep{\fill}}*{4}{c l l l}}
\hline
$x$  & Present method &  Exact value & Error\\
\hline
$0.0$& $1.00000000$&$1.0000000$&$0.00e+00$\\
$0.1$& $0.99833720$&$0.9983358$&$1.40e-06$\\
$0.5$& $0.95984209$&$0.9598391$&$2.99e-06$\\
$1.0$& $0.85505959$&$0.8550576$&$1.99e-06$\\
$5.0$& $0.11082019$&$0.1108198$&$3.89e-07$\\
$6.0$& $0.04373912$&$0.0437380$&$1.12e-06$\\
$6.8$& $0.00417826$&$0.0041678$&$1.05e-05$\\
$6.896$& $0.00003610$&$0.0000360$&$9.79e-08$\\
\hline
\end{tabular*}\label{Tab.STLaneEmden.values3}
\end{table}
%-----------------------------------------------------------------------
\begin{table}
\caption{Comparison of $y(x)$ values of standard Lane-Emden equation, for the present method and exact values given by Horedt \cite{Horedt}, for m=4}
\begin{tabular*}{\columnwidth}{@{\extracolsep{\fill}}*{4}{c l l l}}
\hline
$x$  & Present method &  Exact value & Error\\
\hline
$0.0$ & $1.0000000$&$1.0000000$&$0.00e+00$\\
$0.1$ & $0.9985876$&$0.9983367$&$2.51e-04$\\
$0.2$ & $0.9936339$&$0.9933862$&$2.48e-04$\\
$0.5$ & $0.9605160$&$0.9603109$&$2.05e-04$\\
$1.0$ & $0.8610072$&$0.8608138$&$1.93e-04$\\
$5.0$ & $0.2358368$&$0.2359227$&$8.59e-05$\\
$10.0$& $0.0596105$&$0.0596727$&$6.22e-05$\\
$14.0$& $0.0083058$&$0.0083305$&$2.47e-05$\\
$14.9$& $0.0005759$&$0.0005764$&$4.59e-07$\\
\hline
\end{tabular*}\label{Tab.STLaneEmden.values4}
\end{table}
%-----------------------------------------------------------------------
\begin{table}\tiny
\caption{Coefficients of the Hermite functions of the standard Lane-Emden equations for $m= 2, 3$ and $4$ respectively}
\begin{tabular*}{\columnwidth}{@{\extracolsep{\fill}}*{8}{c r r r c r r r}}
\hline
$i$ & $$ & \multicolumn{1}{c@{\hspace{3pt}}@{\hspace{3pt}}}{$a_i$} & \multicolumn{1}{c@{\hspace{3pt}}||@{\hspace{3pt}}}{$$} & $i$ & $$ & \multicolumn{1}{c@{\hspace{3pt}}@{\hspace{3pt}}}{$a_i$} & $$\\
\hline
$$  & \multicolumn{1}{c@{\hspace{3pt}}@{\hspace{3pt}}}{$m=2$} & \multicolumn{1}{c@{\hspace{3pt}}@{\hspace{3pt}}}{$m=3$} & \multicolumn{1}{c@{\hspace{3pt}}||@{\hspace{3pt}}}{$m=4$} & $$  & \multicolumn{1}{c@{\hspace{3pt}}@{\hspace{3pt}}}{$m=2$} & \multicolumn{1}{c@{\hspace{3pt}}@{\hspace{3pt}}}{$m=3$} & \multicolumn{1}{c@{\hspace{3pt}}@{\hspace{3pt}}}{$m=4$} \\
\hline
$0$ & $-5.2841135322e-01$ & $-4.4099373672e-01$ & \multicolumn{1}{r@{\hspace{3pt}}||@{\hspace{3pt}}}{$-3.8511246127e-01$} & $13$ & $-$ & $-2.4548768173e-02$ & $-$ \\
$1$ & $-2.0672313847e-01$ & $-1.5728415017e-01$ & \multicolumn{1}{r@{\hspace{3pt}}||@{\hspace{3pt}}}{$1.1585058556e-01$} & $14$ & $-$ & $-1.7916420281e-02$ & $-$ \\
$2$ & $-2.1013493211e-01$ & $-1.7607131187e-01$ & \multicolumn{1}{r@{\hspace{3pt}}||@{\hspace{3pt}}}{$-1.6576622713e-01$} & $15$ & $-$ & $-1.0200227258e-02$ & $-$ \\
$3$ & $-1.2898718939e-01$ & $-1.1378421470e-01$ & \multicolumn{1}{r@{\hspace{3pt}}||@{\hspace{3pt}}}{$9.2854106306e-03$} & $16$ & $-$ & $-6.5268030714e-03$ & $-$ \\
$4$ & $-1.3634530855e-01$ & $-1.2995159559e-01$ & \multicolumn{1}{r@{\hspace{3pt}}||@{\hspace{3pt}}}{$-7.1551541010e-02$} & $17$ & $-$ & $-2.7962896018e-03$ & $-$ \\
$5$ & $-8.7619773995e-02$ & $-9.6296863459e-02$ & \multicolumn{1}{r@{\hspace{3pt}}||@{\hspace{3pt}}}{$-9.8809929827e-03$} & $18$ & $-$ & $-1.5765572392e-03$ & $-$ \\
$6$ & $-7.2750465809e-02$ & $-9.8373526479e-02$ & \multicolumn{1}{r@{\hspace{3pt}}||@{\hspace{3pt}}}{$-4.8372346356e-02$} & $19$ & $-$ & $-3.7895054857e-04$ & $-$ \\
$7$ & $-3.9156883681e-02$ & $-7.9430021072e-02$ & \multicolumn{1}{r@{\hspace{3pt}}||@{\hspace{3pt}}}{$-9.4556001733e-03$} & $20$ & $-$ & $-2.4542997154e-04$ & $-$ \\
$8$ & $-2.6813942695e-02$ & $-7.8340439572e-02$ & \multicolumn{1}{r@{\hspace{3pt}}||@{\hspace{3pt}}}{$-2.6810185942e-02$} &  $$ & $$ & $$ & $$ \\
$9$ & $-9.5249929620e-03$ & $-6.2915940155e-02$ & \multicolumn{1}{r@{\hspace{3pt}}||@{\hspace{3pt}}}{$-6.6016540826e-03$} &  $$ & $$ & $$ & $$ \\
$10$ & $-4.1991804282e-03$ & $-5.7157720774e-02$ & \multicolumn{1}{r@{\hspace{3pt}}||@{\hspace{3pt}}}{$-1.1910053277e-02$} & $$ & $$ & $$ & $$ \\
$11$ & \multicolumn{1}{c@{\hspace{3pt}}@{\hspace{3pt}}}{$-$} & $-4.3579589433e-02$ & \multicolumn{1}{r@{\hspace{3pt}}||@{\hspace{3pt}}}{$-1.1402951223e-03$} & $$ & $$ & $$ & $$ \\
$12$ & \multicolumn{1}{c@{\hspace{3pt}}@{\hspace{3pt}}}{$-$} & $-3.6177724390e-02$ & \multicolumn{1}{r@{\hspace{3pt}}||@{\hspace{3pt}}}{$-3.1134305650e-03$} & $$ & $$ & $$ & $$ \\
\hline
\end{tabular*}\label{Tab.STLanemden.coef}
\end{table}
%--------------------------------------------------------------------------------

\begin{table}
\caption{Comparison of $y(x)$ , between present method and series solution given by Wazwaz \cite{Wazwaz.Appl} for isothermal gas sphere equation}
\begin{tabular*}{\columnwidth}{@{\extracolsep{\fill}}*{4}{c r r r}}
\hline
$x$  & Present method &  Wazwaz & Error\\
\hline
$0.0$& $ 0.0000000000$&$ 0.0000000000$&$0.00e+00$ \\
$0.1$& $-0.0016664188$&$-0.0016658339$&$5.85e-07$ \\
$0.2$& $-0.0066539713$&$-0.0066533671$&$6.04e-07$ \\
$0.5$& $-0.0411545150$&$-0.0411539568$&$5.58e-07$ \\
$1.0$& $-0.1588281737$&$-0.1588273537$&$8.20e-07$ \\
$1.5$& $-0.3380198308$&$-0.3380131103$&$6.72e-06$ \\
$2.0$& $-0.5598233120$&$-0.5599626601$&$1.39e-04$ \\
$2.5$& $-0.8063410846$&$-0.8100196713$&$3.68e-03$ \\
%$3.0$& $-1.0633353142$&$-1.0999784903$&$3.66e-02$ \\
\hline
\end{tabular*}\label{Tab.ISOThemalgasValues}
\end{table}
%---------------------------------------------------------------------------
\begin{table}
\caption{Comparison of $y(x)$ , between present method and series solution given by Wazwaz \cite{Wazwaz.Appl} for Example. 3}
\begin{tabular*}{\columnwidth}{@{\extracolsep{\fill}}*{4}{c r r r}}
\hline
$x$  & Present method &  Wazwaz & Error\\
\hline
$0.0$& $1.0000000000$&$1.0000000000$ & $0.00e+00$\\
$0.1$& $0.9981138095$&$0.9980428414$&$7.10e-05$\\
$0.2$& $0.9922758837$&$0.9921894348$&$8.64e-05$\\
$0.5$& $0.9520376245$&$0.9519611019$&$7.65e-05$\\
$1.0$& $0.8183047481$&$0.8182516669$&$5.31e-05$\\
$1.5$& $0.6254886192$&$0.6258916077$&$4.03e-04$\\
$2.0$& $0.4066479695$&$0.4136691039$&$7.02e-03$\\
%$2.5$& $0.1932991104$&$0.2499373217$&$5.66e-02$\\
 \hline
\end{tabular*}\label{Tab.Sinh.y.Values}
\end{table}
%---------------------------------------------------------------------------
\begin{table}
\caption{Comparison of $y(x)$ , between present method and series solution given by Wazwaz \cite{Wazwaz.Appl} for Example. 4}
\begin{tabular*}{\columnwidth}{@{\extracolsep{\fill}}*{4}{c r r r}}
\hline
$x$  & Present method &  Wazwaz & Error\\
\hline
$0.0$ &$1.0000000000$&$1.0000000000$&$0.00e+00$\\
$0.1$& $0.9986051425$&$0.9985979358$&$7.21e-06$\\
$0.2$& $0.9944062706$&$0.9943962733$&$1.00e-05$\\
$0.5$& $0.9651881683$&$0.9651777886$&$1.04e-05$\\
$1.0$& $0.8636881301$&$0.8636811027$&$7.03e-06$\\
$1.5$& $0.7050524103$&$0.7050419247$&$1.05e-05$\\
$2.0$& $0.5064687568$&$0.5063720330$&$9.67e-05$\\
%$2.5$& $0.2924023053$&$0.2912315016$&$1.17e-03$\\
%$3.0$& $0.0910422798$&$0.0821061035$&$8.94e-03$\\
\hline
\end{tabular*}\label{Tab.Sin.y.Values}
\end{table}
%---------------------------------------------------------------------------
\begin{table}
\caption{Comparison of $y(x)$ , between present method and exact solution for Example. 5}
\begin{tabular*}{\columnwidth}{@{\extracolsep{\fill}}*{4}{c r r r}}
\hline
$x$  & Present method &  Exact value & Error\\
\hline
$0.00$& $ 0.0000000000$&$ 0.0000000000$&$0.00e+00$\\
$0.01$& $-0.0001970587$&$-0.0001999900$&$2.93e-06$\\
$0.10$& $-0.0198967225$&$-0.0199006617$&$3.94e-06$\\
$0.50$& $-0.4462840851$&$-0.4462871026$&$3.02e-06$\\
$1.00$& $-1.3862934297$&$-1.3862943611$&$9.31e-07$\\
$2.00$& $-3.2188763248$&$-3.2188758249$&$5.00e-07$\\
$3.00$& $-4.6051709964$&$-4.6051701860$&$8.10e-07$\\
$4.00$& $-5.6664274573$&$-5.6664266881$&$7.69e-07$\\
$5.00$& $-6.5161937402$&$-6.5161930760$&$6.64e-07$\\
$6.00$& $-7.2218363729$&$-7.2218358253$&$5.48e-07$\\
$7.00$& $-7.8240461812$&$-7.8240460109$&$1.70e-07$\\
$8.00$& $-8.3487734467$&$-8.3487745398$&$1.09e-06$\\
$9.00$& $-8.8134506165$&$-8.8134384945$&$1.21e-05$\\
$10.00$&$-9.2302027821$&$-9.2302410337$&$3.83e-05$\\
\hline
\end{tabular*}\label{Tab.ee.y.Values}
\end{table}
%-------------------------------------------------------------
\begin{table}
\caption{Comparison of $y(x)$ , between present method and exact solution for Example. 6}
\begin{tabular*}{\columnwidth}{@{\extracolsep{\fill}}*{4}{c r r r}}
\hline
$x$  & Present method &  Exact value & Error \\
\hline
$0.00$& $1.0000000000$&$1.0000000000$&$0.00e+00$\\
$0.01$& $1.0000999826$&$1.0001000050$&$2.24e-08$\\
$0.02$& $1.0004000642$&$1.0004000800$&$1.58e-08$\\
$0.05$& $1.0025031064$&$1.0025031276$&$2.12e-08$\\
$0.10$& $1.0100501492$&$1.0100501671$&$1.79e-08$\\
$0.20$& $1.0408107527$&$1.0408107742$&$2.15e-08$\\
$0.50$& $1.2840253862$&$1.2840254167$&$3.05e-08$\\
$0.70$& $1.6323161777$&$1.6323162200$&$4.23e-08$\\
$0.80$& $1.8964808279$&$1.8964808793$&$5.14e-08$\\
$0.90$& $2.2479078937$&$2.2479079867$&$9.29e-08$\\
$1.00$& $2.7182819166$&$2.7182818285$&$8.81e-08$\\
\hline
\end{tabular*}\label{Tab.4yln.y.Values}
\end{table}
%-------------------------------------------------------------
\begin{table}
\caption{Comparison of $y(x)$ , between present method and exact solution for Example. 7}
\begin{tabular*}{\columnwidth}{@{\extracolsep{\fill}}*{4}{c r r r}}
\hline
$x$  & Present method &  Exact value & Error \\
\hline
$0.00$& $1.0000000000$&$1.0000000000$&$0.00e+00$\\
$0.01$& $1.0000999826$&$1.0001000050$&$2.24e-08$\\
$0.02$& $1.0004000642$&$1.0004000800$&$1.58e-08$\\
$0.05$& $1.0025031065$&$1.0025031276$&$2.12e-08$\\
$0.10$& $1.0100501493$&$1.0100501671$&$1.78e-08$\\
$0.20$& $1.0408107533$&$1.0408107742$&$2.09e-08$\\
$0.50$& $1.2840253904$&$1.2840254167$&$2.62e-08$\\
$0.70$& $1.6323161872$&$1.6323162200$&$3.27e-08$\\
$0.80$& $1.8964808414$&$1.8964808793$&$3.79e-08$\\
$0.90$& $2.2479079319$&$2.2479079867$&$5.48e-08$\\
$1.00$& $2.7182818260$&$2.7182818285$&$2.51e-09$\\
\hline
\end{tabular*}\label{Tab.ex2.y.Values}
\end{table}
%---------------------------------------------------------------------------
\begin{table}
\caption{Comparison of $y(x)$ , between present method and exact solution for Example. 8}
\begin{tabular*}{\columnwidth}{@{\extracolsep{\fill}}*{4}{c r r r}}
\hline
$x$  & Present method &  Exact value & Error\\
\hline
$0.00$&   $ 0.0000000000$&   $0.0000000000$&$0.00e+00$\\
$0.01$&   $-0.0000009321$&  $-0.0000009900$&$5.79e-08$\\
$0.10$&   $-0.0009008409$&  $-0.0009000000$&$8.41e-07$\\
$0.50$&   $-0.0625021958$&  $-0.0625000000$&$2.20e-06$\\
$1.00$&   $-0.0000008284$&   $0.0000000000$&$8.28e-07$\\
$2.00$&   $ 8.0000001732$&   $8.0000000000$&$1.73e-07$\\
$3.00$&   $54.0000002074$&  $54.0000000000$&$2.07e-07$\\
$4.00$&  $192.0000000368$& $192.0000000000$&$3.68e-08$\\
$5.00$&  $499.9999998091$& $500.0000000000$&$1.91e-07$\\
$6.00$& $1079.9999995264$&$1080.0000000000$&$4.74e-07$\\
$7.00$& $2058.0000004141$&$2058.0000000000$&$4.14e-07$\\
$8.00$& $3584.0000093640$&$3584.0000000000$&$9.36e-06$\\
$9.00$& $5831.9999560359$&$5832.0000000000$&$4.40e-05$\\
$10.00$&$8999.9996608001$&$9000.0000000000$&$3.39e-04$\\
\hline
\end{tabular*}\label{Tab.nh1.y.Values}
\end{table}
%------------------------------------------------------------------
\begin{table}
\caption{Comparison of $y(x)$ , between present method and exact solution for Example. 9}
\begin{tabular*}{\columnwidth}{@{\extracolsep{\fill}}*{4}{c r r r}}
\hline
$x$  & Present method &  Exact value & Error\\
\hline
$0.00$&     $0.0000000000$&   $0.0000000000$&$0.00e+00$        \\
$0.01$&     $0.0000995275$&   $0.0001010000$&$1.47e-06$ \\
$0.10$&     $0.0109981790$&   $0.0110000000$&$1.82e-06$ \\
$0.50$&     $0.3749985918$&   $0.3750000000$&$1.41e-06$ \\
$1.00$&     $1.9999987524$&   $2.0000000000$&$1.25e-06$ \\
$2.00$&    $11.9999993068$&  $12.0000000000$&$6.93e-07$ \\
$3.00$&    $35.9999999242$&  $36.0000000000$&$7.58e-08$ \\
$4.00$&    $80.0000003071$&  $80.0000000000$&$3.07e-07$ \\
$5.00$&   $150.0000003207$& $150.0000000000$&$3.21e-07$ \\
$6.00$&   $252.0000000974$& $252.0000000000$&$9.74e-08$ \\
$7.00$&   $391.9999997951$& $392.0000000000$&$2.05e-07$ \\
$8.00$&   $575.9999992644$& $576.0000000000$&$7.36e-07$ \\
$9.00$&   $810.0000046092$& $810.0000000000$&$4.61e-06$ \\
$10.00$& $1099.9999875537$&$1100.0000000000$&$1.24e-05$ \\
\hline
\end{tabular*}\label{Tab.nh2.y.Values}
\end{table}
%------------------------------------------------------------------
\clearpage
%--------------------------------------------------------------------
%\clearpage{}
%------------------------------------------------------------------------
%\clearpage{}
%------------------------------------------------------------------------
%\clearpage{}
%------------------------------------------------------------------------
%\clearpage{}
%------------------------------------------------------------------------
%\clearpage{}
%------------------------------------------------------------------------
\clearpage{}
%--------------------------------------------------------------------------
%\clearpage{}
%------------------------------------------------------------------------
%\clearpage{}
%-------------------------------------------------------------------------
%\clearpage{}
%-------------------------------------------------------------------------
%\clearpage{}
%-------------------------------------------------------------------------
%\clearpage{}

%-------------------------------------------------------------------------
%\clearpage{}
%--------------------------------------------------------------------------
\end{document}